\documentclass[superscriptaddress, prd, aps,amsmath,amssymb,showpacs,showkeys, onecolumn]{revtex4-2}
\usepackage[dvips]{graphicx,color}
\usepackage{subfig}
\usepackage{times}
\usepackage{xcolor}
\usepackage[%
  colorlinks=true,
  urlcolor=blue,
  linkcolor=red,
  citecolor=blue
]{hyperref}
\usepackage{orcidlink}

\begin{document}
\title{Corrected Thermodynamics and Stability of Magnetic charged AdS Black Holes surrounded by Quintessence }

\author{Dhruba Jyoti Gogoi\orcidlink{0000-0002-4776-8506}}
\email[Email: ]{moloydhruba@yahoo.in}

\affiliation{Department of Physics, Moran College, Moranhat, Charaideo 785670, Assam, India.}
\affiliation{Theoretical Physics Division, Centre for Atmospheric Studies, Dibrugarh University, Dibrugarh
786004, Assam, India.}

\author{Yassine Sekhmani\orcidlink{0000-0001-7448-4579}}
\email[Email: ]{sekhmaniyassine@gmail.com}
\affiliation{Center for Theoretical Physics, Khazar University, 41 Mehseti Street, Baku, AZ1096, Azerbaijan.}
\affiliation{Centre for Research Impact \& Outcome, Chitkara University Institute of Engineering and Technology, Chitkara University, Rajpura, 140401, Punjab, India.}

\author{Shyamalee Bora }
\email[Email: ]{swamaleebora@gmail.com}

\affiliation{Department of Physics, Tezpur University, Tezpur, 784028, Assam, India.}

\author{ Javlon Rayimbaev\orcidlink{0000-0001-9293-1838}}
\email[Email: ]{javlon@astrin.uz}
\affiliation{Institute of Fundamental and Applied Research, National Research University TIIAME, Kori Niyoziy 39, Tashkent 100000, Uzbekistan}
\affiliation{University of Tashkent for Applied Sciences, Gavhar Str. 1, Tashkent 100149, Uzbekistan }
\affiliation{Urgench State University, Kh. Alimdjan str. 14, Urgench 220100, Uzbekistan}
\affiliation{ Shahrisabz State Pedagogical Institute, Shahrisabz Str. 10, Shahrisabz 181301, Uzbekistan}
\affiliation{Tashkent State Technical University, Tashkent 100095, Uzbekistan}

\author{Jyatsnasree Bora \orcidlink{0000-0001-9751-5614}}
\email[Email: ]{jyatnasree.borah@gmail.com}

\affiliation{Department of Physics, D.H.S.K. College, Dibrugarh, 786001, Assam, India}

\author{Ratbay Myrzakulov \orcidlink{0000-0002-5274-0815}}
\email[Email: ]{rmyrzakulov@gmail.com }
\affiliation{L. N. Gumilyov Eurasian National University, Astana 010008,
Kazakhstan.}
\affiliation{Ratbay Myrzakulov Eurasian International Centre for Theoretical
Physics, Astana 010009, Kazakhstan.}

\begin{abstract}
In this study, we explore the corrected thermodynamics of non-linear magnetic charged anti-de Sitter (AdS) black holes surrounded by quintessence, incorporating thermal fluctuations and deriving the corrected thermodynamic potentials. We analyze the effects of corrections due to thermal fluctuations on various thermodynamic potentials, including enthalpy, Helmholtz free energy, and Gibbs free energy. Our results show significant impacts on smaller black holes, with first-order corrections destabilizing them, while second-order corrections enhance stability with increasing parameter values. The specific heat analysis further elucidates the stability criteria, indicating that the large black holes ensure stability against phase transitions. However, the thermal fluctuations do not affect the physical limitation points as well as the second-order phase transition points of the black hole. Our findings highlight the intricate role of thermal fluctuations in black hole thermodynamics and their influence on stability, providing deeper insights into the behaviour of black holes under corrected thermodynamic conditions.

\end{abstract}

\keywords{Black Hole Thermodynamics;  Corrected Entropy; Thermal Fluctuations; Hawking Temperature.}

\maketitle
\section{Introduction}\label{sec01}

Black hole thermodynamics plays a crucial role in formulating any meaningful theory of quantum gravity. This study area, where black holes are treated as thermodynamic objects, was first started by J. D. Bekenstein in 1973. Anything that goes inside a black hole is lost forever. This means that the entropy carried by this object is also lost. If black holes have zero entropy, in that case, the total entropy of the universe will decrease, which is a violation of the second law of thermodynamics \cite{sb1}. Therefore, he pointed out that black holes should have entropy \cite{sb1}. Later on, in 1974, S. Hawking showed that the event horizon of a black hole emits thermal radiation, named Hawking radiation \cite{sb2}. Hawking also proved that the surface area of a black hole cannot decrease in any process \cite{li2}. This is analogous to the second law of Thermodynamics, which states that the entropy of a closed thermodynamic system always increases. Hence, the laws of thermodynamics may apply to black holes, and thermodynamic concepts like entropy, pressure, specific heat, etc., can be associated with black holes.

The area law proposed by Bekenstein and Hawking establishes a relation between entropy ($S$) and area ($A$) of the event horizon and is given by $S=\frac{A}{4}$ \cite{m1,li2}. This means the maximum entropy of black holes is proportional to the event horizon's area. However, the size of black holes decreases due to Hawking radiation. This brings out a need for quantum corrections to the maximum entropy of small-sized black holes in the area law due to thermal fluctuations around the equilibrium. This also led to the modification of the Holographic principle. However, it was discovered that quantum fluctuations do not affect the modified Hayward black hole's stability after studying logarithmic corrections to its thermodynamics \cite{m39}. Correction to the entropy of a black hole results in modification in various thermodynamical equations of states \cite{m13,m14,m15,m15a,m16,m17,m18,m19,m20}. The first-order approximation correction to the black hole entropy is the logarithmic correction \cite{li3}. In the articles \cite{m25} and  \cite{m26}, the authors have analyzed the first-order corrections to the dilatonic black hole and the Schwarzchild Beltrami-de sitter black hole.  The study of higher-order corrections to the black hole entropy due to statistical fluctuations has recently gained much attention.

It is well known that black hole thermodynamics is directly affected by the dark energy component of the universe, ``quintessence" being one of them. Hence, how the thermodynamic quantities related to a black hole are affected by the presence of quintessence is widely studied.  For the charged Reissener-Nordström black hole, Chamblin {\it et al.} have investigated the holography, thermodynamics, and fluctuations \cite{m28}. Hawking and Page's analysis suggests that when the temperature hits a critical point, AdS Schwarzschild's black hole experiences a phase shift. The $P-V$ criticality of black holes is now being studied as a consequence \cite{m29}. As a result of thermal fluctuations, the first-order corrections to a dumb hole, the black hole's counterpart, have been examined in Ref. \cite{m36}.
Additionally, Pourhassan {\it et al.} \cite{m37,m38} investigated the thermodynamic effects of statistical thermal fluctuations on a charged dilatonic black Saturn and single spinning Kerr-Ads black hole. Nevertheless, it was discovered that quantum fluctuations did not affect the modified Hayward black hole's stability by analyzing logarithmic thermodynamic adjustments \cite{m39}. The introduction of quantum fluctuations elevated from small thermal fluctuations has occurred recently. Other noteworthy research addressing various aspects of black hole characteristics in various gravity theories can be found in articles \cite{Banerjee1, Banerjee2, Lambiase01,dj, Banerjee3, Banerjee4, Banerjee5}. In addition to these works, there have been several investigations to explore different thermal, optical and perturbative properties associated with black holes in different modified theory of gravity frameworks \cite{mm1,mm2,mm3,mm4,mm5,mm6,mm7,mm8,mm9,mm10,mm11,mm12,mm13,mm14,mm15,mm16, mm17, mm18, mm19, mm20, mm21, mm22, mm23, mm24}. These works will contribute towards our understanding of black hole systems in different modified gravity frameworks. 

Thermodynamics of magnetically charged AdS black holes in an extended phase space considering the coupling of nonlinear electrodynamics are studied in Ref. \cite{kruglov}. In 2018, C. H. Nam derived a non-linear magnetic-charged black hole surrounded by quintessence \cite{nam}. The author has studied its thermodynamic stability. It reported that the black hole may undergo a thermal phase transition, at some critical temperature, between a larger unstable black hole and a smaller stable black hole \cite{nam}. Among the various techniques developed to address the issue of singularity, ordinary black holes are one. The Bardeen black hole is an example of a black hole that does not involve singularity. Its event horizon satisfies the weak energy requirement. This was determined by introducing a tensor representing energy and momentum, understood to be the gravitational field of a non-linear magnetic monopole charge Q. This has led to a growing interest in the geometrical and thermodynamic features of these regular black holes among researchers. Recently, a study on the thermodynamic features of a non-linear magnetic-charged AdS black hole surrounded by quintessence in the background of perfect fluid dark matter is reported in Ref. \cite{ndo}. The corrected thermodynamics of a non-linear magnetic charged AdS black hole surrounded by perfect fluid dark matter has been studied by R. Ndongmo {\it et al.} and showed that small-sized black holes are affected by thermal fluctuations as the second law of thermodynamics is violated by the corrected entropy which in turn leads to a non-linear evolution and entropy decrease. They also showed that the larger black holes are not affected by thermal fluctuations \cite{ndo}. {Apart from these there are several other investigations dealing with corrected thermodynamics of black holes \cite{Jawad:2020ihz, Zhang:2018nep, Pourhassan:2017bip, Jawad:2017mwt, Pourhassan:2016qoz, Pourhassan:2020yei, Chatterjee:2020iuf, Pourhassan:2022cvn, Nadeem-ul-islam:2020dhd, Fatima:2024gji, Pourhassan:2018scc}.  }

With these motivations of the above-discussed work, we will study the higher-order entropy corrections for a non-linear magnetic-charged AdS black hole surrounded by a quintessence field. Further, we shall analyze other thermal properties of this black hole system and derive corrected expressions for its enthalpy, Helmholtz free energy, thermodynamic volume, specific heat, internal energy, and Gibbs' free energy. Our study focuses on modifying various thermodynamic parameters of non-linear magnetic charged AdS black holes due to thermal fluctuations around the equilibrium point.

Throughout this paper, we stick to the metric signature $(-,+,+,+)$ and consider the geometrical units (fundamental constants) $c=G=\hbar=k_{B}=1$. After this brief introduction, the rest of the paper is arranged as follows. In section \ref{sec2}, we briefly overview the geometry of non-linear magnetic-charged AdS black holes. In section \ref{sec3}, we briefly study the Hawking temperature and basic thermodynamic variables. The effects of thermal fluctuations on entropy have been studied in \ref{sec4}. The impacts of thermal fluctuations on thermodynamic potentials have been investigated in section \ref{sec5}. In section \ref{sec6}, we study the stability of non-linear magnetic-charged AdS black holes under the effect of thermal fluctuations. Finally, we summarize our results in section \ref{sec7}.

\section{Non-linear magnetic-charged AdS black hole surrounded by quintessence} \label{sec2}

This section will discuss the physics of non-linear magnetic charged AdS black holes. For this, we will consider the action corresponding to the Einstein gravity coupled to a non-linear electromagnetic field in the four-dimensional AdS space-time and surrounded by a quintessence field as, ~\cite{li2012galactic,xu2018kerr,xu2019perfect,sadeghi2020universal,salazar1987duality,novello2000singularities,nam2020higher}
\begin{eqnarray}\label{act}
\begin{array}{r c l}
S&=&\int d^4x\sqrt{-g}\left[\frac{c^4}{16\pi G}(R-2\Lambda)-(\mathcal{L}_\textmd{charge}-\mathcal{L}_\textmd{quint})\right]
\end{array}
\end{eqnarray}
In this action, $R$ represents the scalar curvature, $g$ is the determinant of the metric tensor $g_{\mu\nu}$, $G$ is the Newton gravity constant, and $c$ is the light speed. Also, $\Lambda$ is the cosmological constant, and $\mathcal{L}_\textmd{charge}$ is the non-linear electrodynamic term. This  $\mathcal{L}_\textmd{charge}$ is a function of the invariant quantity $F_{\mu\nu}F^{\mu\nu}/4\equiv F$, and $F_{\mu\nu}$ is the Faraday tensor of electromagnetic field expressed as $F_{\mu\nu}=\partial_\mu A_\nu-\partial_\nu A_\mu$. Where $A_\nu$ is the gauge potential of the electromagnetic field. In Eq. \eqref{act}, the non-linear electrodynamic term $\mathcal{L}_\textmd{charge}$ and the quintessence term $\mathcal{L}_\textmd{quint}$ can be defined respectively as follows \cite{nam2018on,nam2018non,chen2020optical,ma2021shadow,sadeghi2020ads,ghosh2018lovelock,bohmer2015interacting},
\begin{equation}\label{L_F}
\mathcal{L}_\textmd{charge}=\frac{3M}{|Q|^3}\frac{(2Q^2F)^{3/2}}{\left[1+(2Q^2F)^{3/4}\right]^2}\,
\end{equation}
\begin{equation}
\mathcal{L}_\textmd{quint}=-\frac{1}{2}(\nabla\phi)^2-V(\phi).
\end{equation}
Here, the parameters $M$ and $Q$ are associated with the system's mass and magnetic charge. Also, $\phi$ is the quintessential scalar field, and $V(\phi)$ represents the potential. Here, it can be noted that the magnetic charge $Q$ is defined as~\cite{nam2018on} 
\begin{equation}\label{sphere}
\frac{1}{4\pi}\int_{S_2^\infty}\textbf{\textit{F}}=Q,
\end{equation}
where $S_2^\infty$ is a two-sphere at the infinity. 

Upon application of the variational principle from Eq. (\ref{act}), with respect to the inverse of the metric $g^{\mu\nu}$, one can have, 
\begin{eqnarray}
\label{S1}    0&=&\frac{1}{\sqrt{-g}}\frac{\delta S}{\delta g^{\mu\nu}}\\ 
\label{S2} &=&\frac{1}{\sqrt{-g}}\left\{\frac{c^4}{16\pi G}\int d^4x \frac{\delta (\sqrt{-g}(R-2\Lambda))}{\delta g^{\mu\nu}} -\left(\int d^4x \frac{\delta (\sqrt{-g}\mathcal{L}_\textmd{charge})}{\delta g^{\mu\nu}}- \int d^4x \frac{\delta (\sqrt{-g}\mathcal{L}_\textmd{quint})}{\delta g^{\mu\nu}}\right)\right\}.
\end{eqnarray}

This will lead to,
\begin{equation}\label{Guv-2}
\begin{array}{r c l}
G_{\mu\nu}+\Lambda g_{\mu\nu}&=&\frac{8\pi G}{c^4}\left(T_{\mu\nu}(\textmd{charge})-\mathcal{T}_{\mu\nu}\right),
\end{array}
\end{equation}
here the different energy-momentum tensors are defined as \cite{nam2018non,gonzalez2008exact,bohmer2015interacting},
\begin{eqnarray}
\label{Tuv-3}   T_{\mu\nu}(\textmd{charge})= \frac{2\delta (\sqrt{-g}\mathcal{L}_\textmd{charge})}{\delta g^{\mu\nu}}=-\frac{\partial \mathcal{L}_\textmd{charge}}{\partial F}F_\mu^\beta F_{\nu\beta}+\mathcal{L}_\textmd{charge}g_{\mu\nu},\\ 
\label{Tuv-2}  \mathcal{T}_{\mu\nu} = \frac{2\delta (\sqrt{-g}\mathcal{L}_\textmd{quint})}{\delta g^{\mu\nu}} =\left[\nabla_\mu\phi\nabla_\nu\phi-\frac{1}{2}g_{\mu\nu}\left((\nabla\phi)^2+2V(\phi)\right)\right].
\end{eqnarray}

In this work we will consider the unit system where the Newton gravity constant $G$, light speed $c$ and the number $\pi$ are such that $\frac{4\pi G}{c^4}=1$. With this unit system, we finally arrive at the Einstein-Maxwell equations of motion as follows:
\begin{eqnarray}
\begin{array}{r c l}
\label{Guv}   G^\nu_\mu+\Lambda \delta^\nu_\mu&=&2\left(\frac{\partial\mathcal{L}_\textmd{charge}}{\partial F}F_{\mu\rho}F^{\nu\rho}-\delta^\nu_\mu\mathcal{L}_\textmd{charge}-\mathcal{T}^\nu_\mu\right),
\end{array}
\end{eqnarray}
\begin{eqnarray}
\label{Max1}  \nabla_\mu\left(\frac{\partial\mathcal{L}_\textmd{charge}}{\partial F}F^{\nu\mu}\right)&=0,\\
\label{Max2}  \nabla_\mu*F^{\nu\mu}&=0.
\end{eqnarray}

To find a spherically symmetric AdS black hole solution of the mass $M$ and the magnetic charge $Q$ in the quintessence, the metric has to be written with ansatz~\cite{nam2018on,xu2018kerr,rizwan2020coexistent}
\begin{eqnarray}\label{metric}
\begin{array}{r c l}
ds^2&=&-e^\nu dt^2+e^\lambda dr^2+r^2(d\theta^2+\sin^2\theta d\phi^2)\\ 
&=&-f(r)dt^2+\frac{1}{f(r)}dr^2+r^2(d\theta^2+\sin^2\theta d\phi^2),\\
&\textnormal{and}&\ \ f(r)=1-\frac{2m(r)}{r}-\frac{\Lambda}{3}r^2.
\end{array}
\end{eqnarray}

One may note here that the term $Q$ (as defined in Eq.\eqref{sphere}) is the integral constant that should be used to integrate Eqs. (\ref{Max1}) and (\ref{Max2}) to find the expression of $F_{\mu\nu}$. Also, the term $M$ will allow us to find the expression of $m(r)$. Now, taking into account Eqs. (\ref{Max1}), (\ref{Max2}) and (\ref{sphere}), we can reach the following expression ~\cite{nam2018on},
\begin{equation}\label{F}
F=\frac{Q^2}{2r^4}.
\end{equation}
Thus from Eq. (\ref{L_F}), the non-linear electrodynamic term can be rewritten as
\begin{equation}
\mathcal{L}_\textmd{charge}=\frac{3MQ^3}{(r^3+Q^3)^2}.
\end{equation}

Now using the time component of Eq. (\ref{Guv})
\begin{eqnarray}
\begin{array}{r c l}
\label{Guv1}     G^t_t+\Lambda \delta^t_t&=&2\left(\frac{\partial\mathcal{L}_\textmd{charge}}{\partial F}F_{t\rho}F^{t\rho}-\delta^t_t\mathcal{L}_\textmd{charge}-\mathcal{T}^t_t\right).
\end{array}
\end{eqnarray}
Moreover, as the time components of energy-momentum for quintessence are related to their energy density, so we have ~\cite{zhang2021regular,Kiselev2003,ndo},
\begin{eqnarray}\label{Tuv1} 
\mathcal{T}^t_t&=&-\frac{3\epsilon c_q}{2r^{3(\epsilon+1)}},
\end{eqnarray}
here $c_q$ and $\epsilon$ are the quintessence parameters.

Therefore, from Eq. (\ref{Guv1}), one can arrive at the following expression 
\begin{equation}\label{Guv2}
G_t^t+\Lambda=-2\mathcal{L}_\textmd{charge}+\frac{3\epsilon c_q}{r^{3(\epsilon+1)}}.
\end{equation}

To solve this Eq. (\ref{Guv2}), first we will find the component $G_{tt}$ of the Einstein tensor as
\begin{equation}\label{Guv3}
G_{tt}=e^\nu\left[\frac{1}{r^2}-e^{-\lambda}\left(\frac{1}{r^2}-\frac{\lambda'}{r}\right)\right].
\end{equation}
Developing it, we get
\begin{eqnarray*}
G_{tt}&=&f(r)\left[\frac{1}{r^2}-f(r)\left(\frac{1}{r^2}-\frac{1}{r}\left(-\frac{f'(r)}{f(r)}\right)\right)\right] \\
&=& f(r)\left[\frac{1}{r^2}-f(r)\left(\frac{1}{r^2}-\frac{1}{rf(r)}\left[2\left(\frac{1}{r}\frac{dm(r)}{dr}
-\frac{m(r)}{r^2}+\frac{\Lambda}{3}r^2\right)\right]\right)\right]\\
&=&f(r)\left(\frac{2}{r^2}\frac{dm(r)}{dr}+\Lambda\right),
\end{eqnarray*}
with $\lambda=-\ln f(r)$.

Thus one can arrive at the component $G_t^t$ of Eq. (\ref{Guv3}) as
\begin{equation}
G_t^t=g^{tt}G_{tt}=(-f(r))^{-1}f(r)\left(\frac{2}{r^2}\frac{dm(r)}{dr}+\Lambda\right),
\end{equation}
or,
\begin{equation}\label{Guv4}
G_t^t=-\frac{2}{r^2}\frac{dm(r)}{dr}-\Lambda.
\end{equation}

Now, using this Eq. (\ref{Guv4}) in Eq. (\ref{Guv2}), we get 
\[
\frac{dm(r)}{dr}=\frac{3MQ^3r^2}{(r^3+Q^3)^2}-\frac{3\epsilon c_q}{2r^{3\epsilon+1}}.
\]
Upon integration, we obtain
\begin{equation}
m(r)=-\frac{MQ^3}{r^3+Q^3}+\frac{c_q}{2r^{3\epsilon}}+C_{0}.
\end{equation}
here the integral constant $C_{0}$ can be obtained using the boundary condition~\cite{nam2018on,zhang2021regular, ndo} 
\begin{eqnarray*}
M&=&\lim\limits_{r\rightarrow \infty}\left\{m(r)-\frac{c_q}{2r^{3\epsilon}}\right\}\\
&=&\lim\limits_{r\rightarrow \infty}\left\{-\frac{MQ^3}{r^3+Q^3}+C_{0}\right\},
\end{eqnarray*}
therefore
\begin{equation}
C_{0}=M.
\end{equation}

The expression for the mass $m(r)$ and the function $f(r)$ can be given as
\begin{equation}
m(r)=\frac{Mr^3}{r^3+Q^3}+\frac{c_q}{2r^{3\epsilon}},
\end{equation}
\begin{equation}
f(r)=1-\frac{2Mr^2}{r^3+Q^3}-\frac{c_q}{r^{3\epsilon+1}}-\frac{\Lambda}{3}r^2.
\end{equation}

Finally, we reach the following expression for the spherically symmetric solution for the action \eqref{act} as
\begin{equation}
\label{metric1}
ds^2=-f(r)dt^2+\frac{1}{f(r)}dr^2+r^2(d\theta^2+\sin^2\theta d\phi^2),
\end{equation}
\begin{equation}
\label{eq27}
\textnormal{with}\ f(r)=1-\frac{2Mr^2}{r^3+Q^3}-\frac{c_q}{r^{3\epsilon+1}}-\frac{\Lambda}{3}r^2.
\end{equation}
For the quintessence field, we have chosen $\epsilon=-2/3$.
When we put quintessence parameter $c_q=0$ and the cosmological constant $\Lambda=0$ into Eq. (\ref{metric1}), we retrieve the metric of a non-linear magnetic-charged black hole.

\begin{figure*}[t!]
      	\centering{
      	\includegraphics[scale=0.85]{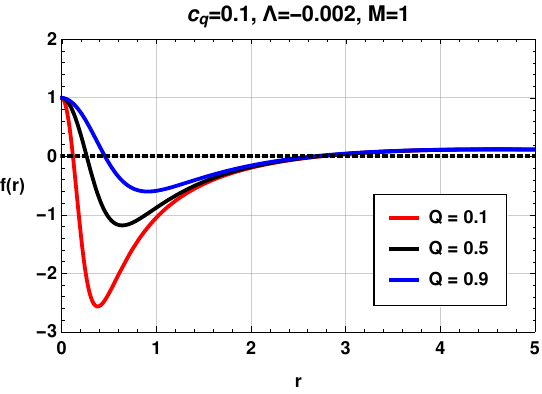} \hspace{5mm}
       \includegraphics[scale=0.85]{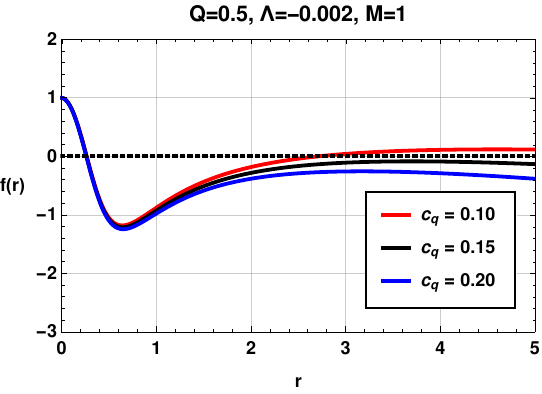}\\
       }
      	\caption{Behaviour of the metric function of the black hole.}
      	\label{horizon01}
      \end{figure*}

{We have investigated the horizon structure of the black hole metric \eqref{metric1} in Fig. \ref{horizon01}. One may observe that on the first panel of Fig. \ref{horizon01}, the charge parameter $Q$ only affects the event horizon of the black hole and it doesn't impact the cosmological horizon of the black hole. The black hole, as seen from the behaviour of the metric function, is a regular one for $Q\ne0$. On the second panel of Fig. \ref{horizon01}, we observe that the quintessence parameter $c_q$ has no impact on the event horizon of the black hole. But it affects the cosmological horizon significantly. We observe that the cosmological horizon appears only for smaller values of the quintessence parameter $c_q$ and as $c_q$ increases, the black hole only has an event horizon and no cosmological horizon. }

\section{Temperature and Basic Thermodynamical Properties}\label{sec3}

In this section, the thermodynamical properties of non-linear magnetic charged black holes will be studied, for which we will proceed through the discussion about the event horizon. By solving the equation $f(r)=0$ at the event horizon, we get the horizon radius $r_+$, which will give the volume of the event horizon as:
\begin{equation}\label{eq28}
    V = \frac{4 \pi  r_+^3}{3}.
\end{equation}
Using the given lapse function $f(r)$ i.e., Eq.(\ref{eq27}), we can derive the Hawking temperature $T_{H}$ as follows,
\begin{equation}\label{5}
T_{H}=\frac{f^\prime(r)}{4\pi}\Bigg|_{r\ = \ r_{+}}=-\frac{-Q^3 r_+ c_q+2 r_+^4 c_q+2 Q^3+\Lambda  r_+^5-r_+^3}{4 \pi  Q^3 r_++4 \pi  r_+^4},
\end{equation}
where prime ``$\prime$" represents the differentiation with respect to $r$, and $r_{+}$ denotes the horizon radius of a black hole. However, we can observe that 
\begin{equation}\label{5a}
\lim_{(\Lambda, c_q, Q)\to(0,0, 0)} T_{H} = \frac{1}{8\pi M}
\end{equation}
which is similar to the Hawking temperature calculated for the Schwarzschild black hole (SBH).
\begin{figure*}[t!]
      	\centering{
      	\includegraphics[scale=0.65]{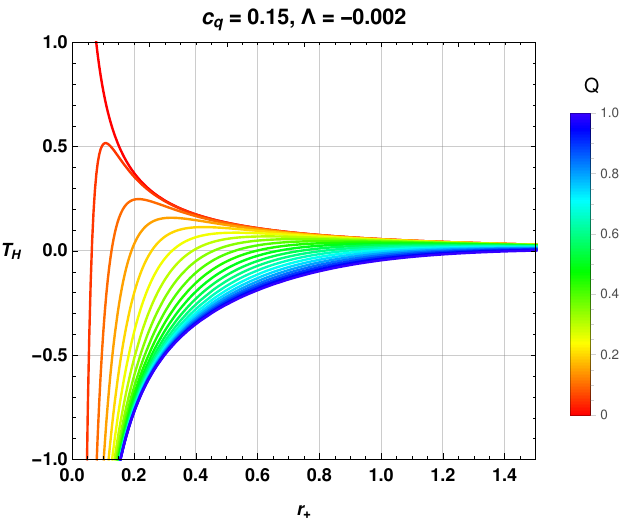} \hspace{5mm}
       \includegraphics[scale=0.65]{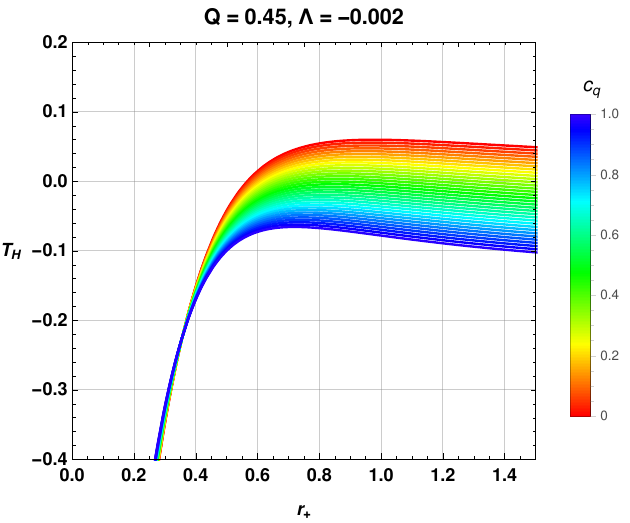}\\
       \includegraphics[scale=0.65]{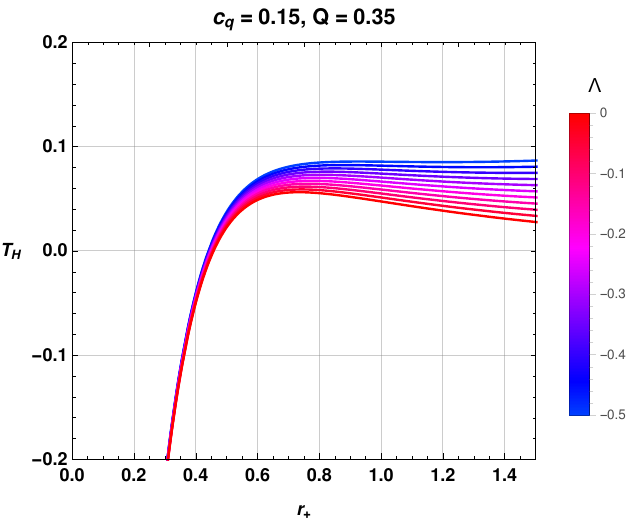}
       
       }
      	\caption{Variation of the Hawking temperature with the black hole horizon radius $r_+$.}
      	\label{figT01}
      \end{figure*}
We have shown the variation of the black hole temperature in Fig. \ref{figT01}. One can see from the first panel of Fig. \ref{figT01} that the magnetic charge has a noticeable impact on the temperature of the black hole. For a small amount of charge $Q$, the temperature is negative only for very small black holes, and as $r_+$ rises, the temperature becomes positive and slowly decreases as $r_+$ increases. With an increase in the value of $Q$, $T_H$ becomes negative even for comparatively large black holes. The quintessence parameter $c_q$ also impacts the temperature significantly. In the presence of the quintessence field, the temperature of the black hole decreases, as shown in the second panel of Fig. \ref{figT01}. Finally, the impact of the cosmological constant, which mimics pressure in the black hole system on the temperature, is shown in the third panel of Fig. \ref{figT01}. It also has a similar impact on $T_H$ as demonstrated by the quintessence field.

{One may observe from Fig. \ref{figT01} that the temperature of this black hole becomes negative in some certain regions. From the horizon structure analysis shown in Fig. \ref{horizon01}, one may note that the black hole event horizon is affected by $Q$ only and the cosmological horizon is affected by $c_q$ only for a fixed value of $M$ and $\Lambda$. Moreover, $Q=0$ results in a singular black hole. From the analysis of temperature, we see that for the singular black hole, the temperature becomes positive for any horizon radius of the black hole. But in the case of a singular black hole {\it i.e.,} for $Q\ne0$, the temperature of smaller black holes becomes negative. On the other hand, a large value of $c_q$ also results in a negative temperature of comparatively larger black holes. In the context of this magnetically charged regular AdS black hole surrounded by a quintessence field, the concept of negative temperature introduces fascinating quantum and classical dynamics. A negative temperature may represent systems that are hotter than any positive temperature system \cite{Norte:2024crb}. AdS spacetime, with its distinctive negatively curved geometry, plays a significant role in defining the thermodynamic properties of black holes, such as temperature and pressure \cite{Toledo:2019amt}. In this scenario, negative temperature reflects an inverted population of energy states within the black hole, where higher energy states dominate over lower ones. This inversion can occur due to the long-range interactions inherent in AdS spacetime, coupled with the intense gravitational forces near the event horizon, creating a situation similar to quantum systems that exhibit negative temperatures.

The presence of magnetic charge further influences the behaviour of the black hole, modifying the surrounding electromagnetic field and adding complexity to its thermodynamics. In a state of negative temperature, the magnetic charge could enhance repulsive forces that work against gravitational collapse. This results in an outward pressure, counteracting the inward pull of gravity and stabilizing the black hole in an equilibrium state. The thermodynamic inversion suggested by negative temperature in such a charged black hole implies a higher-than-expected thermal energy, aligning with quantum mechanical behaviour seen in other systems with negative temperatures.

Surrounding this black hole is a quintessence field, a form of dark energy that exerts negative pressure. Quintessence is typically linked to the accelerated expansion of the universe, but in the vicinity of the black hole, it could interact with the black hole’s thermodynamics in a way that enables or enhances the negative temperature state. The quintessence field's contribution to the overall negative pressure complements the outward quantum pressure produced by the black hole’s negative temperature. This dual force could help to stabilize the black hole, countering gravitational forces that would otherwise drive collapse.

In this framework, the interaction between quantum gravitational effects, AdS spacetime, magnetic charge, and the quintessence field creates a unique thermodynamic system. The negative temperature suggests that near the event horizon, quantum effects might impose energy limits that result in a state where outward pressure prevents collapse. These forces help maintain the regularity of the black hole, avoiding the formation of singularities. The existence of negative temperature in this case points to a deeper quantum-gravitational interplay, where the balance between quantum effects and gravity stabilizes the black hole and presents novel thermodynamic behaviours \cite{Norte:2024crb}.}

The corresponding Bekenstein entropy can be calculated using the four fundamental laws of Black hole thermodynamics, which leads us to the uncorrected entropy:
\begin{equation}\label{eq31}
    S _0= \frac{\pi  \left(r_+^3-2 Q^3\right)}{r_+}.
\end{equation}
The enthalpy energy of the system is calculated using the standard formula:
\begin{equation}\label{eq32}
     H = \int T_HdS_0.
\end{equation}
This gives the enthalpy energy of the non-linear magnetic charged AdS black hole system as:
\begin{equation}\label{eq33}
    H = \frac{1}{2} \left(-\frac{Q^3 c_q}{r_+}-r_+^2 c_q+\frac{Q^3}{r_+^2}-\frac{\Lambda  r_+^3}{3}+r_+\right).
\end{equation}
The expression for pressure is given as:
\begin{equation}\label{eq34}
P=-\frac{\Lambda}{8\pi}.
\end{equation}
The above derived quantities are used to obtain the expressions for the other system properties, such as internal energy $(U)$, Helmholtz Free energy $F$, and specific heat $C$ for the system of Non-linear magnetic charged AdS black holes. 
We use the familiar expression $U=H-PV$ to find the internal energy and using the expressions from \eqref{eq34}, \eqref{eq33} and \eqref{eq28} the internal energy is:
\begin{equation}
    U = -\frac{\left(Q^3+r_+^3\right) \left(r_+ c_q-1\right)}{2 r_+^2}.
\end{equation}
In the following sections, we shall discuss the effect of small stable fluctuations near equilibrium of the thermodynamical properties of the non-linear magnetic charged AdS black hole system. 

\section{Thermodynamic fluctuations: the second-order corrections to entropy}\label{sec4}

In this section, we calculate the effect of thermal fluctuations on the entropy of the black hole. The seminal paper by Hawking and Page \cite{hawp} shows that black holes in asymptotically curved space-time can be described using a canonical ensemble. Based on this, we consider the magnetically charged AdS black holes system as a canonical ensemble consisting of $N$ particles with an energy spectrum $E_{n}$. The statistical partition function of the system can be written as:
\begin{equation}
    Z = \int_0^\infty  dE  \rho (E) e^{-\bar{\beta}_{\kappa} E},
\end{equation}
where $\bar{\beta}_{\kappa}$ is the inverse temperature in units of the Boltzmann constant, and $\rho (E)$ is the canonical density of the system with average energy $E$. Using the partition function $Z$ and Laplace inversion, the density of states can be calculated as:

\begin{equation}
    \rho (E) = \frac{1}{2 \pi i} \int^{\bar{\beta}_{0\kappa}+
i\infty}_{\bar{\beta}_{0\kappa} - i\infty} d \bar{\beta}_{\kappa}  e^{S(\bar{\beta}_{\kappa})},
\end{equation}

where
\begin{equation}
    S = \bar{\beta}_{\kappa} E + \log Z.
\end{equation}

The entropy around the equilibrium temperature $\bar{\beta}_{0\kappa}$ is obtained by eliminating all thermal fluctuations. However, in the presence of thermal fluctuations, the corrected entropy can be expressed by a Taylor expansion around $\bar{\beta}_{0\kappa}$ as:

\begin{equation}
    S = S_0 + \frac{1}{2}(\bar{\beta}_{\kappa} - \bar{\beta}_{0\kappa})^2 \left(\frac{\partial^2 S(\bar{\beta}_{\kappa})}{\partial \bar{\beta}_{\kappa}^2 }\right)_{\bar{\beta}_{\kappa} = \bar{\beta}_{0\kappa}} + \frac{1}{6}(\bar{\beta}_{\kappa} - \bar{\beta}_{0\kappa})^3 \left(\frac{\partial^3 S(\bar{\beta}_{\kappa})}{\partial \bar{\beta}_{\kappa}^3 }\right)_{\bar{\beta}_{\kappa} = \bar{\beta}_{0\kappa}} + \cdots,
\end{equation}

where the dots denote higher-order corrections.

We should note that the first derivative of the entropy for $\bar{\beta}_{\kappa}$ vanishes at the equilibrium temperature. Consequently, the density of states can be expressed as:
\begin{equation}
    \rho (E) = \frac{e^{S_0}}{2 \pi i} \int^{\bar{\beta}_{0\kappa} + i\infty}_{\bar{\beta}_{0\kappa} - i\infty} d \bar{\beta}_{\kappa} \, \exp \left( \frac{(\bar{\beta}_{\kappa} - \bar{\beta}_{0\kappa})^2}{2} \left(\frac{\partial^2 S(\bar{\beta}_{\kappa})}{\partial \bar{\beta}_{\kappa}^2 }\right)_{\bar{\beta}_{\kappa} = \bar{\beta}_{0\kappa}} + \frac{(\bar{\beta}_{\kappa} - \bar{\beta}_{0\kappa})^3}{6} \left(\frac{\partial^3 S(\bar{\beta}_{\kappa})}{\partial \bar{\beta}_{\kappa}^3 }\right)_{\bar{\beta}_{\kappa} = \bar{\beta}_{0\kappa}} + \cdots \right).
\end{equation}

Following the approach in Ref. \cite{More:2004hv}, we obtain:
\begin{equation}
    S = S_0 - \frac{1}{2} \log{S_{0}T_{h}^{2}} + \frac{f(m,n)}{S_{0}} + \cdots,
\end{equation}

where $f(m,n)$ is considered a constant. A more general expression for the corrected entropy is given by \cite{PF}:
\begin{equation}
    S = S_0 - \frac{\beta_1}{2} \log(S_0 T_{H}^2) + \frac{\beta_2}{S_0} + \cdots,
\end{equation}

where the parameters $\beta_1$ and $\beta_2$ are introduced to track the first-order and second-order corrected terms. {The first order corrected term for a charged AdS black hole {\it i.e.,} $\beta_1$ was initially discussed in Ref. \cite{mm13}.} When $\beta_1 \rightarrow 0$ and $\beta_2 \rightarrow 0$, the original results are recovered, and $\beta_1 = 1$ and $\beta_2 = 0$ yield the usual corrections \cite{More:2004hv,SPR}. Therefore, the first-order correction is logarithmic, while the second-order correction is proportional to the inverse of the original entropy $S_0$. These corrections can be considered quantum corrections to the black hole. For large black holes, these corrections can be neglected. However, as the black hole decreases in size due to Hawking radiation, the quantum fluctuations in the black hole's geometry increase. Thus, thermal fluctuations significantly modify the thermodynamics of black holes \cite{Mandal:2023ahb}, becoming more important as the black holes reduce in size. {Black holes with smaller areas, approaching dimensions on the order of Planck's length, may require non-perturbative corrections, such as exponential corrections, to accurately describe their behaviour \cite{Pourhassan:2020yei,Pourhassan:2022cvn}. As the black hole shrinks due to Hawking radiation, these non-perturbative corrections become dominant near the Planck scale \cite{Pourhassan:2020yei,Pourhassan:2022cvn}. In this regime, perturbative corrections are insufficient, and the exponential terms arise when microstate counting is performed for quantum states confined to the black hole's horizon \cite{Chatterjee:2020iuf}. However, in our analysis, the results remain valid only above the Planck scale, where we have considered only perturbative correction terms.}
\begin{figure*}[t!]
      	\centering{
      	\includegraphics[scale=0.65]{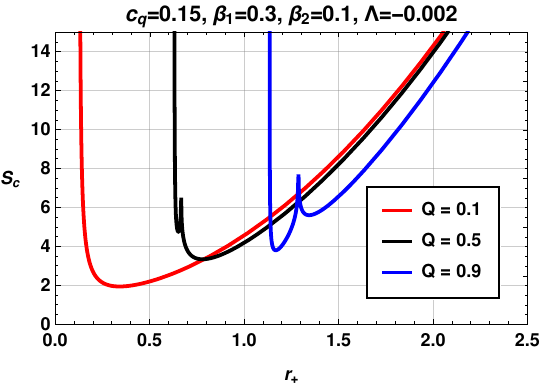}
       \hspace{5mm}
            \includegraphics[scale=0.65]{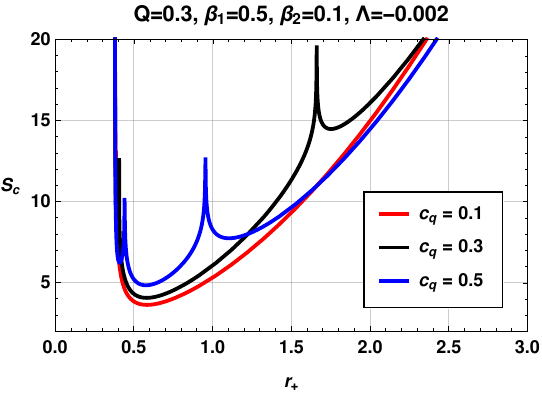}\\
            \includegraphics[scale=0.65]{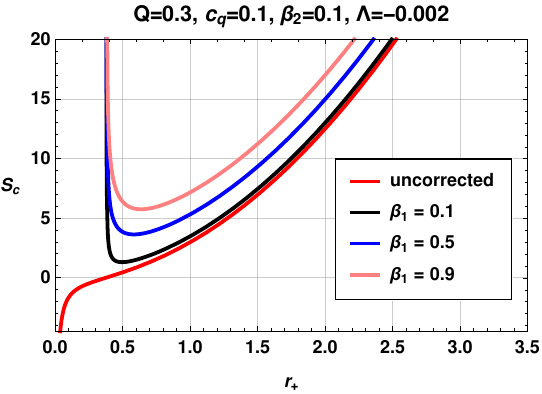}
            \hspace{5mm}
             \includegraphics[scale=0.65]{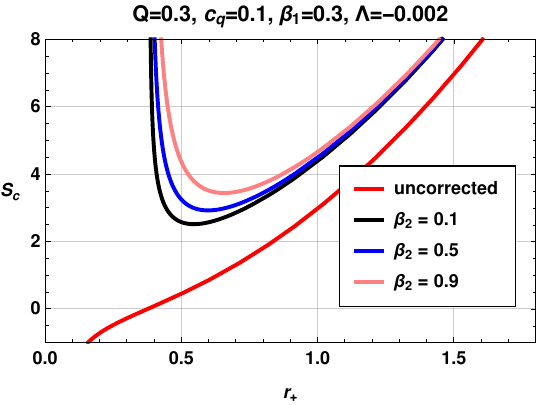}
            }
      	\caption{Variation of entropy with the black hole horizon radius.}
      	\label{figSc}
      \end{figure*}
      
Up to the second-order correction, the explicit form of the entropy for this black hole is given by
\begin{equation}
    S_c = \frac{\pi  \left(r_+^3-2 Q^3\right)}{r_+}-\beta _1 \log \left(\frac{\pi  \left(r_+^3-2 Q^3\right) \left(-Q^3 r_+ c_q+2 r_+^4 c_q+2 Q^3+\Lambda  r_+^5-r_+^3\right){}^2}{r_+ \left(4 \pi  Q^3 r_++4 \pi  r_+^4\right){}^2}\right)+\frac{\beta _2 r_+}{\pi  \left(r_+^3-2 Q^3\right)}.
\end{equation}

The variation of the corrected entropy of the black hole is shown in Fig. \ref{figSc}. Here, we have demonstrated the variations concerning different parameters of this study. With a comparatively small magnetic charge Q and zero quintessence parameter $c_q$, the corrected entropy is a smooth increasing function of $r_{+}$. It is shown in the first panel of the figure. However, we observe that the nature of increasing entropy for large charge Q undergoes a sudden rise in entropy. These peaks shift towards a large horizontal radius for large Q values. In the second panel, considering $Q=0.3$, we investigate the entropy variation for a set of quintessence parameter $c_q$. These variations in entropy also suggest regular increasing values with some significant peaks. However, for $c_q=0.1$, no such significant peaks are observed. In the last two panels of this figure, corrected and uncorrected entropy variations are shown for the first-order and second-order corrected terms $\beta_1$ and $\beta_2$, taking the rest parameters to have fixed values. All these variations only take positive values for all the cases considered.

In the next section, we will analyze the thermodynamic quantities of the black hole with the corrected entropy.

\section{Impact of Thermal fluctuations: the Second order corrected thermodynamic potentials}\label{sec5}

This section aims to evaluate various thermodynamic variables for the magnetically charged AdS black hole in the presence of thermodynamic fluctuations. As mentioned earlier, we shall consider the second-order entropy corrections here. Using the second-order corrected entropy, it is possible to derive corrected enthalpy energy $H_c$ as given by
\begin{equation}\label{29}
H_{c}=\int T_{H}dS_{c}+ \int V_c dP,
\end{equation}
where $V_{c}$ is the corrected volume. However, as the second term on the RHS of the above equation does not contribute to a constant value of $\lambda$, using the expressions of $T_H$ and $S_c$ in the above equation, we can have
Corrected enthalpy
\begin{multline}
    H_c = -\frac{1}{144 \pi ^2}\left[\frac{72 \pi  \beta _1 \left(3 Q^3 c_q+r_+ \left(\Lambda  Q^3+3 r_+\right)\right)}{Q^3+r_+^3}+\frac{12 \beta _2 r_+ \left(3 c_q+2 \Lambda  r_+\right)}{r_+^3-2 Q^3}+\frac{72 \pi  \left(\pi  Q^3 c_q-3 \beta _1\right)}{r_+}\right.\\\left.-\frac{12 \log \left(r_+^3-2 Q^3\right) \left(3 \pi  \beta _1 Q^3 c_q-\beta _2\right)}{Q^3}-\frac{36 \log \left(r_+\right) \left(\beta _2+\pi  \beta _1 Q^3 c_q\right)}{Q^3}\right.\\\left.+\frac{2^{2/3} \log \left(2 Q^2+2^{2/3} Q r_++\sqrt[3]{2} r_+^2\right) \left(3\ 2^{2/3} \beta _2 c_q+6\ 2^{2/3} \pi  \beta _1 \Lambda  Q^3+4 \beta _2 \Lambda  Q\right)}{Q^2}\right.\\\left.-\frac{2\ 2^{2/3} \log \left(2 Q-2^{2/3} r_+\right) \left(3\ 2^{2/3} \beta _2 c_q+6\ 2^{2/3} \pi  \beta _1 \Lambda  Q^3+4 \beta _2 \Lambda  Q\right)}{Q^2}\right.\\\left.+\frac{2\ 2^{2/3} \sqrt{3} \tan ^{-1}\left(\frac{Q+2^{2/3} r_+}{\sqrt{3} Q}\right) \left(3\ 2^{2/3} \beta _2 c_q+6\ 2^{2/3} \pi  \beta _1 \Lambda  Q^3-4 \beta _2 \Lambda  Q\right)}{Q^2}\right.\\\left.+72 \pi ^2 r_+^2 c_q-\frac{72 \pi ^2 Q^3}{r_+^2}-72\pi r_+ \left(2 \beta_1 \Lambda +\pi \right)+24 \pi^2 \Lambda r_+^3 \right].
    \end{multline}
\begin{figure*}[t!]
      	\centering{
      	\includegraphics[scale=0.65]{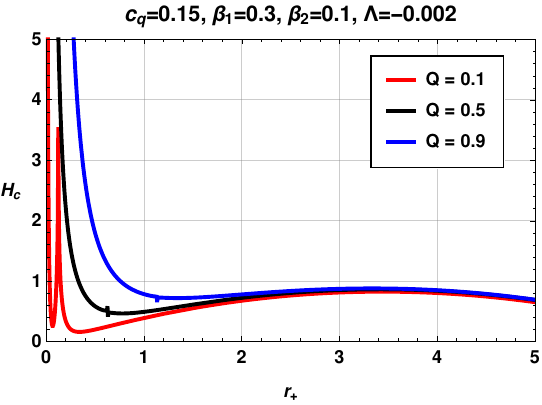} \hspace{5mm}
            \includegraphics[scale=0.65]{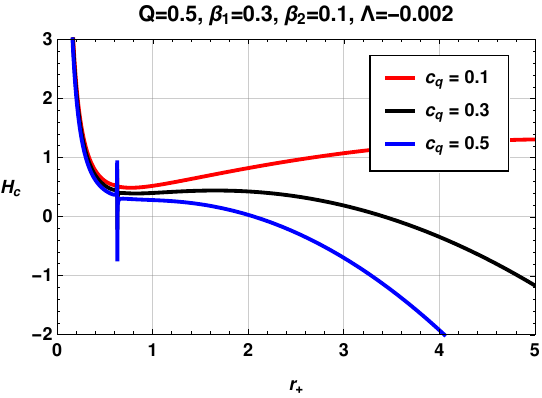}\\
            \includegraphics[scale=0.65]{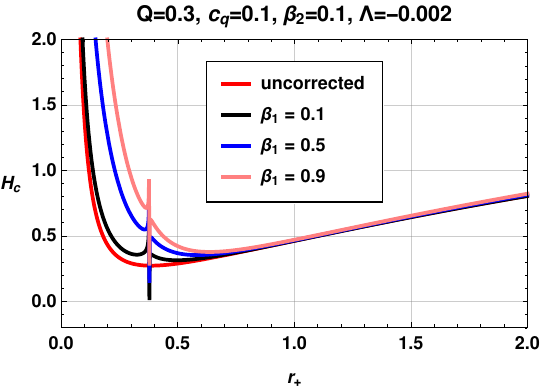}
            \hspace{5mm}
             \includegraphics[scale=0.65]{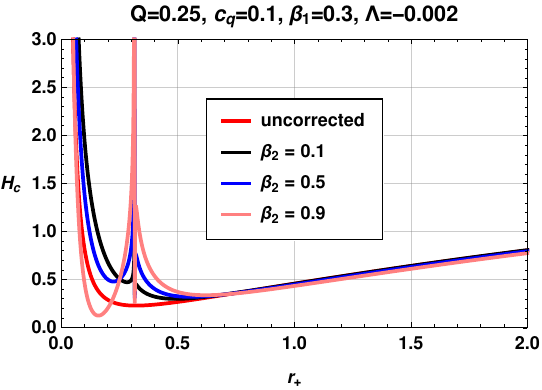}
            }
      	\caption{Variation of enthalpy with the black hole horizon radius.}
      	\label{figHc}
      \end{figure*}
Graphically, the corrected enthalpy $H_c$ is shown in Fig. \ref{figHc} for different values of the model parameters. An increase in the charge parameter $Q$ increases the enthalpy of the black hole. The variation is significant for smaller black holes, while the variation is less significant for larger ones. In the case of the parameter $c_q$, we observe a decrease in the enthalpy as $c_q$ increases. The variations are negligible for smaller black holes and significant for larger ones. It shows that the quintessence parameter $c_q$ signatures are prominent only for large black holes.

One can see that the behaviour of enthalpy is different in the presence of thermal fluctuations. For smaller black holes, thermal fluctuation corrections of both first and second order significantly increase the enthalpy. In comparison, the variations are not very significant for larger black holes. Interestingly, for larger black holes, the presence of thermal fluctuations results in a slight decrease in the enthalpy value. A discontinuity in the graphs is present due to the thermal fluctuations.

Using the corrected enthalpy, one can find an expression for the corrected volume given by
\begin{equation}
    V_c = \frac{dH_c}{dP} \Bigg|_{S_c = constant}
\end{equation}
Using the expressions for $H_c$ in the above relation,
the explicit form of corrected volume is found to be
\begin{multline}
    V_c = \frac{1}{18 \pi }\left[\frac{72 \pi  \beta _1 Q^3 r_+}{Q^3+r_+^3}+\frac{24 \beta _2 r_+^2}{r_+^3-2 Q^3}-\frac{4\ 2^{2/3} \left(2 \beta _2+3\ 2^{2/3} \pi  \beta _1 Q^2\right) \log \left(2 Q-2^{2/3} r_+\right)}{Q}\right.\\\left.+\frac{2\ 2^{2/3} \left(2 \beta _2+3\ 2^{2/3} \pi  \beta _1 Q^2\right) \log \left(2 Q^2+2^{2/3} Q r_++\sqrt[3]{2} r_+^2\right)}{Q}\right.\\\left.+\frac{4\ 2^{2/3} \sqrt{3} \left(3\ 2^{2/3} \pi  \beta _1 Q^2-2 \beta _2\right) \tan ^{-1}\left(\frac{Q+2^{2/3} r_+}{\sqrt{3} Q}\right)}{Q}-144 \pi  \beta _1 r_++24 \pi ^2 r_+^3\right].
\end{multline}
The corrected Helmholtz free energy can be calculated by using the standard relation
\begin{equation}
    F_c = - \int S_c dT_H - \int P dV_c.
\end{equation}
Inserting the corresponding corrected parameters,
\begin{multline}
    F_c = \frac{1}{72 \pi ^2}\left[-\frac{36 \pi ^2 Q^3}{r_+^2}+18 \pi ^2 \Lambda  r_+^3+18 \left(7 \pi  c_q \beta _1+\frac{\beta _2}{Q^3}\right) \log \left(r_+\right)-72 \pi  \Lambda  r_+ \beta _1+72 \pi  \log \left(Q^3+r_+^3\right) c_q \beta _1\right.\\\left.-72 \pi  \log \left(\left(2-r_+ c_q\right) Q^3+r_+^3 \left(\Lambda  r_+^2+2 c_q r_+-1\right)\right) c_q \beta _1\right.\\\left.+\frac{18 \pi  \log \left(\frac{\left(r_+^3-2 Q^3\right) \left(\left(2-r_+ c_q\right) Q^3+r_+^3 \left(\Lambda  r_+^2+2 c_q r_+-1\right)\right){}^2}{16 \pi  r_+^3 \left(Q^3+r_+^3\right){}^2}\right) \left(-\Lambda  r_+^5+r_+^3+Q^3 \left(3 r_+ c_q-2\right)\right) \beta _1}{r_+ \left(Q^3+r_+^3\right)}\right.\\\left.+\frac{108 \pi  \beta _1}{r_+}+18 \pi  r_+ \left(4 \Lambda  \beta _1+\pi \right)+\frac{12 \Lambda  r_+ \left(3 \pi  \left(r_+^3-2 Q^3\right) \beta _1 Q^3+r_+ \left(Q^3+r_+^3\right) \beta _2\right)}{-2 Q^6-r_+^3 Q^3+r_+^6}\right.\\\left.+\frac{2^{2/3} \log \left(2 Q-2^{2/3} r_+\right) \left(6\ 2^{2/3} \pi  \Lambda  \beta _1 Q^3+4 \Lambda  \beta _2 Q+3\ 2^{2/3} c_q \beta _2\right)}{Q^2}\right.\\\left.+\frac{6 \left(9 \pi ^2 \left(\Lambda  Q^3+3 r_+-3 r_+^2 c_q\right) Q^3-6 \pi  \left(3 c_q Q^3+r_+ \left(\Lambda  Q^3+3 r_+\right)\right) \beta _1+\left(\Lambda  r_+^2+3 c_q r_+-3\right) \beta _2\right)}{Q^3+r_+^3}\right.\\\left.-\frac{2^{2/3} \sqrt{3} \tan ^{-1}\left(\frac{Q+2^{2/3} r_+}{\sqrt{3} Q}\right) \left(6\ 2^{2/3} \pi  \Lambda  \beta _1 Q^3-4 \Lambda  \beta _2 Q+3\ 2^{2/3} c_q \beta _2\right)}{Q^2}\right.\\\left.-\frac{\log \left(2 Q^2+2^{2/3} r_+ Q+\sqrt[3]{2} r_+^2\right) \left(6\ 2^{2/3} \pi  \Lambda  \beta _1 Q^3+4 \Lambda  \beta _2 Q+3\ 2^{2/3} c_q \beta _2\right)}{\sqrt[3]{2} Q^2}\right.\\\left.-\frac{6 \log \left(r_+^3-2 Q^3\right) \left(3 \pi  c_q \beta _1 Q^3+\beta _2\right)}{Q^3}-\frac{2\ 2^{2/3} \Lambda  \log \left(2 Q-2^{2/3} r_+\right) \left(3\ 2^{2/3} \pi  \beta _1 Q^2+2 \beta _2\right)}{Q}\right.\\\left.+\frac{2^{2/3} \Lambda  \log \left(2 Q^2+2^{2/3} r_+ Q+\sqrt[3]{2} r_+^2\right) \left(3\ 2^{2/3} \pi  \beta _1 Q^2+2 \beta _2\right)}{Q}\right.\\\left.+\frac{2\ 2^{2/3} \Lambda  \tan ^{-1}\left(\frac{Q+2^{2/3} r_+}{\sqrt{3} Q}\right) \sqrt{3} \left(3\ 2^{2/3} \pi  Q^2 \beta _1-2 \beta _2\right)}{Q}\right],
\end{multline}
which is thermal fluctuation dependent. Without thermal fluctuation correction, it is given by
\begin{equation}
    F = \frac{1}{4} \left(\frac{3 Q^3 \left(-3 r_+^2 c_q+\Lambda  Q^3+3 r_+\right)}{Q^3+r_+^3}-\frac{2 Q^3}{r_+^2}+\frac{\Lambda  r_+^3}{3}+r_+\right)+\frac{\Lambda  r_+^3}{6}.
\end{equation}

\begin{figure*}[t!]
      	\centering{
      	\includegraphics[scale=0.65]{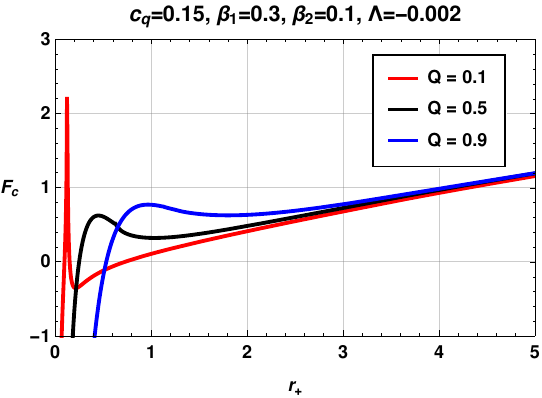} \hspace{5mm}
            \includegraphics[scale=0.65]{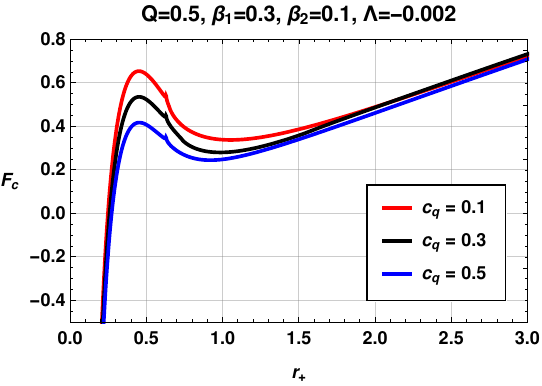}\\
            \includegraphics[scale=0.65]{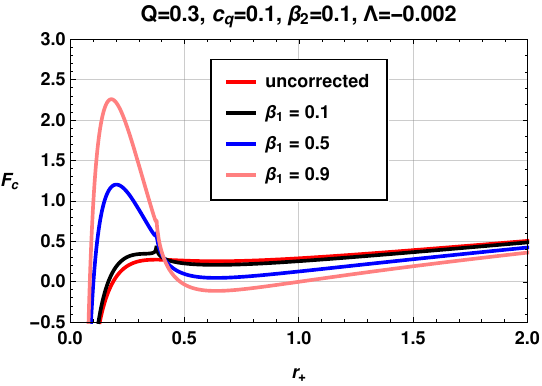}
            \hspace{5mm}
             \includegraphics[scale=0.65]{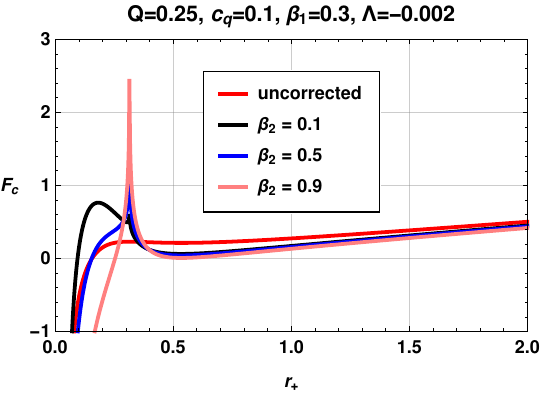}
            }
      	\caption{Variation of Helmholtz free energy with the black hole horizon radius.}
      	\label{figFc}
      \end{figure*}
\begin{figure*}[t!]
      	\centering{
      	\includegraphics[scale=0.65]{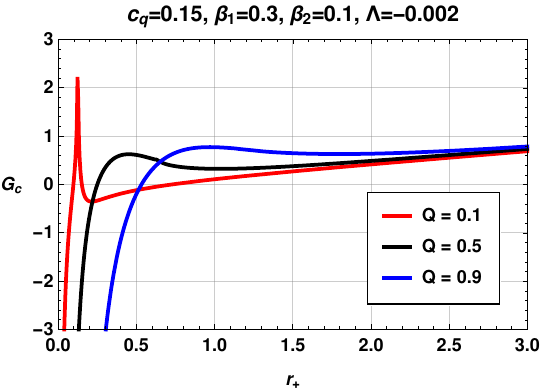} \hspace{5mm}
            \includegraphics[scale=0.65]{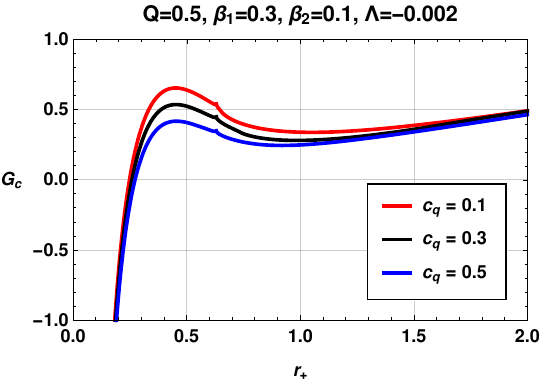}\\
            \includegraphics[scale=0.65]{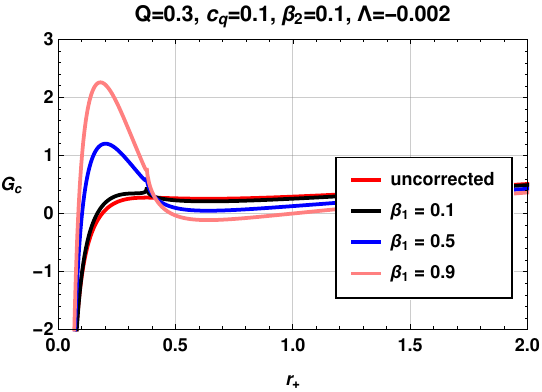}
            \hspace{5mm}
             \includegraphics[scale=0.65]{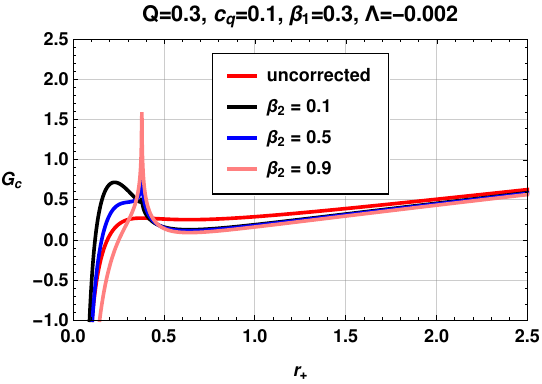}
            }
      	\caption{Variation of Gibbs free energy with the black hole horizon radius.}
      	\label{figGc}
      \end{figure*}
We have shown the corrected Helmholtz free energy variation in Fig. \ref{figFc}. The figures show that both the correction parameters have noticeable impacts on the Helmholtz free energy.
We observe that an increase in the charge $ Q $ shifts the peak of the Helmholtz free energy curve towards larger black holes, and the area under the curve increases gradually. However, an increase in the quintessence parameter $ c_q $ decreases the Helmholtz free energy. For the first-order correction parameter $ \beta_1 $, an increase in its value initially increases the Helmholtz free energy for small-sized black holes, but after crossing a threshold event horizon radius, further increases in $ \beta_1 $ decrease the Helmholtz free energy. Similarly, for the correction parameter $ \beta_2 $, an increase in its value decreases the Helmholtz free energy gradually for small-sized black holes. However, for small values of $ \beta_2 $, the Helmholtz free energy remains greater than that of the corresponding uncorrected black hole. A lower Helmholtz free energy indicates that a black hole is in a more stable thermodynamic state, making states with the lowest Helmholtz free energy the most stable. A decrease in Helmholtz's free energy implies that the system (black hole plus surroundings) is moving towards a more stable equilibrium state.
Furthermore, the Helmholtz free energy represents the maximum amount of work that can be extracted from a black hole system at constant temperature and volume, indicating that any process resulting in a decrease in Helmholtz free energy can, in principle, be harnessed to perform useful work. States with lower Helmholtz free energy are exponentially more probable, and a black hole with a higher Helmholtz free energy is more likely to radiate energy to move toward a more stable state. Therefore, a black hole state with lower Helmholtz free energy achieves a balance where the energy and entropy terms are optimized for stability at a given temperature.

 The corrected internal energy ($U_{c}$) can be obtained from the following formula:
\begin{equation}\label{34}
U_{c}=H_{c}-PV_{c}.
\end{equation}
The explicit form is given by
\begin{multline}
    U_c = \frac{1}{24} \left(-\frac{36 \beta _1 Q^3 c_q}{\pi  \left(Q^3+r_+^3\right)}-\frac{6 \beta _2 r_+ c_q}{\pi ^2 \left(r_+^3-2 Q^3\right)}+\frac{6 \beta _1 c_q \log \left(r_+^3-2 Q^3\right)}{\pi }+\frac{6 \log \left(r_+\right) \left(\beta _2+\pi  \beta _1 Q^3 c_q\right)}{\pi ^2 Q^3}\right.\\\left.-\frac{12 Q^3 c_q}{r_+}+\frac{2 \sqrt[3]{2} \beta _2 c_q \log \left(2 Q-2^{2/3} r_+\right)}{\pi ^2 Q^2}-\frac{\sqrt[3]{2} \beta _2 c_q \log \left(2 Q^2+2^{2/3} Q r_++\sqrt[3]{2} r_+^2\right)}{\pi ^2 Q^2}\right.\\\left.-\frac{2 \sqrt[3]{2} \sqrt{3} \beta _2 c_q \tan ^{-1}\left(\frac{Q+2^{2/3} r_+}{\sqrt{3} Q}\right)}{\pi ^2 Q^2}-12 r_+^2 c_q-\frac{36 \beta _1 r_+^2}{\pi  \left(Q^3+r_+^3\right)}-\frac{2 \beta _2 \log \left(r_+^3-2 Q^3\right)}{\pi ^2 Q^3}+\frac{12 Q^3}{r_+^2}+\frac{36 \beta _1}{\pi  r_+}+12 r_+\right).
\end{multline}

We shall now examine the effects of thermal fluctuations on Gibbs free energy. In thermodynamics, Gibbs free energy quantifies the maximum amount of mechanical work obtainable from a system. The following relation mathematically defines it:
\begin{equation}\label{36}
G_{c}=F_{c}+PV_{c}.
\end{equation}
Using the corrected values of Helmholtz free energy and volume, we can get
\begin{multline}\label{37}
G_{c}= \frac{1}{144 \pi ^2} \Big[-144 \pi  \beta _1 c_q \log \left(Q^3 \left(2-r_+ c_q\right)+r_+^3 \left(2 r_+ c_q+\Lambda  r_+^2-1\right)\right)-\frac{72 \pi  \beta _1 \Lambda  Q^3 r_+}{Q^3+r_+^3}-\frac{24 \beta _2 \Lambda  r_+^2}{r_+^3-2 Q^3}-\frac{72 \pi ^2 Q^3}{r_+^2} \\ +\frac{36 \pi  \beta _1 \left(Q^3 \left(3 r_+ c_q-2\right)-\Lambda  r_+^5+r_+^3\right) \log \left(\frac{\left(r_+^3-2 Q^3\right) \left(Q^3 \left(2-r_+ c_q\right)+r_+^3 \left(2 r_+ c_q+\Lambda  r_+^2-1\right)\right){}^2}{r_+^3 \left(Q^3+r_+^3\right){}^2}\right)}{r_+ \left(Q^3+r_+^3\right)} \\ +\frac{12 \left(-3 \pi  \beta _1 (2+\log (16 \pi )) \left(3 Q^3 c_q+r_+ \left(\Lambda  Q^3+3 r_+\right)\right)+9 \pi ^2 Q^3 \left(-3 r_+^2 c_q+\Lambda  Q^3+3 r_+\right)+\beta _2 \left(3 r_+ c_q+\Lambda  r_+^2-3\right)\right)}{Q^3+r_+^3} \\ +36 \log \left(r_+\right) \left(7 \pi  \beta _1 c_q+\frac{\beta _2}{Q^3}\right)+144 \pi  \beta _1 c_q \log \left(Q^3+r_+^3\right)-\frac{12 \log \left(r_+^3-2 Q^3\right) \left(\beta _2+3 \pi  \beta _1 Q^3 c_q\right)}{Q^3} \\ +\frac{2\ 2^{2/3} \log \left(2 Q-2^{2/3} r_+\right) \left(3\ 2^{2/3} \beta _2 c_q+6\ 2^{2/3} \pi  \beta _1 \Lambda  Q^3+4 \beta _2 \Lambda  Q\right)}{Q^2} +\frac{72 \pi  \beta _1 (3+\log (16 \pi ))}{r_+}+12 \pi ^2 \Lambda  r_+^3 \\ -\frac{2\ 2^{2/3} \sqrt{3} \tan ^{-1}\left(\frac{Q+2^{2/3} r_+}{\sqrt{3} Q}\right) \left(3\ 2^{2/3} \beta _2 c_q+6\ 2^{2/3} \pi  \beta _1 \Lambda  Q^3-4 \beta _2 \Lambda  Q\right)}{Q^2}+\frac{24 \Lambda  r_+ \left(3 \pi  \beta _1 Q^3 \left(r_+^3-2 Q^3\right)+\beta _2 r_+ \left(Q^3+r_+^3\right)\right)}{-2 Q^6-Q^3 r_+^3+r_+^6} \\ +36 \pi  r_+ \left(\beta _1 \Lambda  (4+\log (16 \pi ))+\pi \right) -\frac{2^{2/3} \log \left(2 Q^2+2^{2/3} Q r_++\sqrt[3]{2} r_+^2\right) \left(3\ 2^{2/3} \beta _2 c_q+6\ 2^{2/3} \pi  \beta _1 \Lambda  Q^3+4 \beta _2 \Lambda  Q\right)}{Q^2} \Big].
\end{multline}

This shows that the thermal fluctuations also affect the Gibbs free energy. To visualize the impacts of the first-order and second-order correction parameters, we have plotted the corrected Gibbs free energy in Fig \ref{figGc}. In the first two panels of Fig. \ref{figGc}, we showed the impacts of charge $Q$ and quintessence parameter $c_q$ on the Gibbs free energy of the black hole system. One can see that the corrected Gibbs free energy $G_c$ is negative for smaller black holes. An increase in the value of $Q$ increases the possibility of a negative $G_c$ for the small-sized black holes. After a certain radius of the black hole, $G_c$ becomes positive. As shown in the figure, the peak shifts towards larger black holes as $Q$ increases gradually. However, towards larger values of $r_+$, $Q$ does not have a distinguishable impact. On the other hand, with an increase in the value of the quintessence parameter $c_q$, the peak of the curve does not shift horizontally but instead of this, it decreases gradually. In this case, the impact of $c_q$ is not properly distinguishable for very large and very small black holes as depicted in the figure. 

The impacts of the correction parameters $\beta_1$ and $\beta_2$ are shown in the last two panels of Fig. \ref{figGc}. One can see that both parameters have noticeable impacts on the behaviour of $G_c$. Initially, with an increase in the correction parameter $\beta_1$ ($\beta_2$), $G_c$ increases (decreases) significantly. In this scenario, Gibb's free energy is greater than the corresponding value of the uncorrected black hole. After crossing a threshold value of the event horizon radius, the pattern becomes opposite {\it i.e.,} the uncorrected black hole attains greater Gibbs free energy. This observation suggests that the thermal fluctuations decrease the stability of the small-sized black holes and increase the stability of larger black holes. The negative value of Gibbs free energy for very small-sized black holes indicates that the black hole is thermodynamically stable. This means that small perturbations or fluctuations will not lead to the decay or dramatic changes in the black hole's state. It also represents maximum energy that can be extracted from the thermodynamical system of the black hole.
A negative Gibbs free energy suggests that the entropy increase (disorder) associated with processes involving the black hole is favored, which aligns with the second law of thermodynamics. This is in agreement with the outcomes depicted by the black hole's entropy variations.

\section{ Stability in presence of Thermal fluctuations}\label{sec6}

\begin{figure*}[t!]
      	\centering{
      	\includegraphics[scale=0.65]{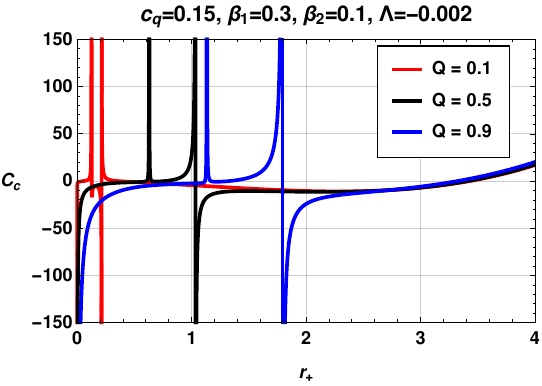} \hspace{5mm}
            \includegraphics[scale=0.65]{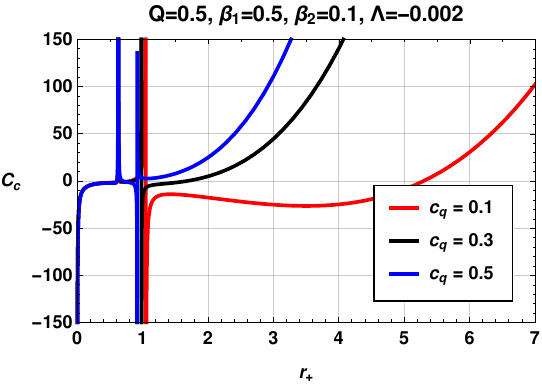}\\
            \includegraphics[scale=0.65]{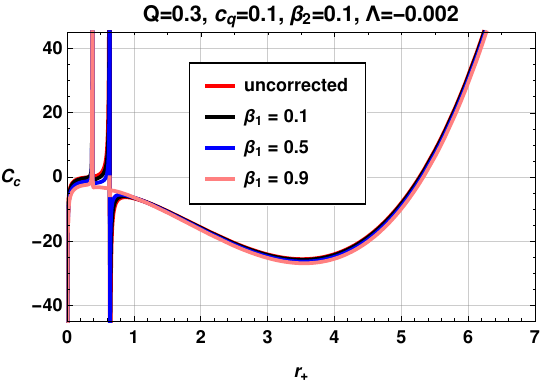}
            \hspace{5mm}
             \includegraphics[scale=0.65]{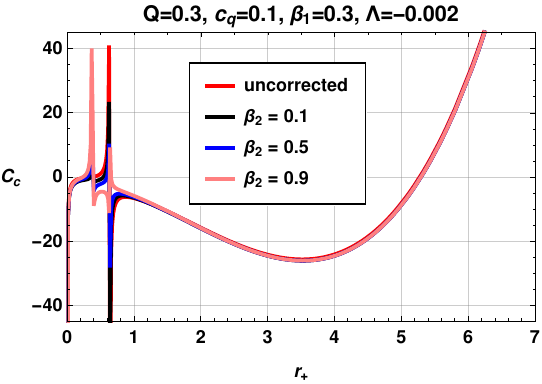}
            }
      	\caption{Variation of specific heat with the black hole horizon radius.}
      	\label{figCc}
      \end{figure*}

To study the stability, we investigate the nature of its specific heat. We can explore whether this black hole undergoes a phase transition from the specific heat behaviour. The positive value of specific heat ensures that the given black hole system is stable against the phase transition, however, the negative value of specific heat concludes the system's instability. By considering thermal fluctuation, we estimate the expression for specific heat, which must subside to the uncorrected specific heat (when fluctuation is switched off, i.e. $\beta=0$). From a classical thermodynamics point of view, the specific heat ($C_{c}$) can be calculated with the help of the following standard formula: 
\begin{equation}\label{38}
C_{c}=T_{H}\frac{dS_{c}}{dT_{H}}.
\end{equation}
Or, in terms of parameter space, the specific heat can be identified in the form of
\begin{multline}
    C_c =  -\frac{1}{\pi  r_+ \left(r_+^3-2 Q^3\right){}^2 \left(Q^3 r_+^3 \left(9 r_+ c_q+4 \Lambda  r_+^2-10\right)-2 Q^6+\Lambda  r_+^8+r_+^6\right)}\Big[2 \Big\lbrace\beta _2 r_+^2 \left(Q^3+r_+^3\right){}^2 \Big(Q^3 \left(2-r_+ c_q\right) \\ +r_+^3 \left(2 r_+ c_q+\Lambda  r_+^2-1\right)\Big)+\pi ^2 \left(2 Q^6+Q^3 r_+^3-r_+^6\right){}^2 \left(Q^3 \left(r_+ c_q-2\right)-r_+^3 \left(2 r_+ c_q+\Lambda  r_+^2-1\right)\right) \\ +\pi  \beta _1 r_+ \left(r_+^3-2 Q^3\right) \left(Q^9 \left(6-r_+ c_q\right)+Q^6 r_+^3 \left(-18 r_+ c_q-7 \Lambda  r_+^2+21\right)+4 Q^3 r_+^6 \left(3 r_+ c_q+\Lambda  r_+^2-3\right)+2 r_+^{10} \left(c_q+\Lambda  r_+\right)\right)\Big\rbrace\Big].
\end{multline}
In absence of the corrections, uncorrected specific heat of the black hole is given by
\begin{equation}
    C_0 = -\frac{2 \pi  \left(Q^3+r_+^3\right){}^2 \left(Q^3 \left(r_+ c_q-2\right)-r_+^3 \left(2 r_+ c_q+\Lambda  r_+^2-1\right)\right)}{Q^3 r_+^4 \left(9 r_+ c_q+4 \Lambda  r_+^2-10\right)-2 Q^6 r_++\Lambda  r_+^9+r_+^7}.
\end{equation}

Concretely, the obtained heat capacity is physically constrained within the pair of corrected parameters $(\beta_1, \beta_2)$ as a corrected term and operates the uncorrected term with the set of $(\beta_1=\beta_2=0)$. It is worth noting that beyond certain limits, the heat capacity brings to such uncorrected heat capacity like Schwarzschild-AdS by the parameter space selection $(\beta_1=0, \beta_2=0, Q=0, c_q=0)$, while to Schwarzschild across the selection $(\beta_1=0, \beta_2=0, Q=0, c_q=0, \Lambda=0)$. As a matter of interest, the examination of the behaviour of the heat capacity $(C_c)$ as a function of the horizon radius $(r_+)$ is based on certain characteristics involving the pertinent Hawking temperature, the mass of the black hole and the corrected entropy. Therefore, it is vital firstly to identify a further feature about the variation in heat capacity involving two specified points, namely the point of physical limitation and the point of divergence representing critical phase transition points of the black hole. More precisely, the subsequent algorithm is useful in the essence of the corrected terms for displaying the specified points
   \begin{eqnarray}
    T_H=\left(\frac{\partial M}{\partial S_c}\right)&=&0\hspace{0.4cm} \text{physical limitation points},\\
         \left(\frac{\partial^2 M}{\partial S_c^2}\right)&=&0\hspace{0.3cm}\text{second order phase transition}.
    \end{eqnarray}

Thus for our black hole scenario, the condition representing the physical limitation points takes place,
\begin{equation}
    \left(2 Q^6+Q^3 r_+^3-r_+^6\right) \left[Q^3 \left(r_+ c_q-2\right)-r_+^3 \left(2 r_+ c_q+\Lambda  r_+^2-1\right)\right]=0.
\end{equation}

The equation governing the second-order phase transitions is given by

\begin{equation}
    \left(2 Q^3-r_+^3\right) \left(Q^3+r_+^3\right) \left[Q^3 \left(r_+ c_q-3\right)+r_+^4 \left(c_q+\Lambda  r_+\right)\right] \left[Q^3 \left(r_+ c_q-2\right)-r_+^3 \left(2 r_+ c_q+\Lambda  r_+^2-1\right)\right]=0.
\end{equation}
From the above two conditions, one can see that both the physical limitation points and second-order phase transition points are independent of the thermal fluctuation parameters $\beta_1$ and $\beta_2$. 

 The other key characteristic to consider is the sign of the heat capacity, which is a crucial indicator of a system's thermal stability. A positive heat capacity implies that the system is thermally stable. This means that when energy is added to the system, its temperature increases in a predictable and controlled manner. Conversely, a negative heat capacity signifies thermal instability. In this scenario, adding energy to the system leads to a decrease in temperature, causing unpredictable and potentially unstable thermal behaviour. Therefore, the sign of the heat capacity serves as a fundamental determinant of whether a system can maintain thermal equilibrium or is prone to thermal fluctuations and instability.~\cite{Sahabandu:2005ma, Cai:2003kt}. Consequently, with these elements in sight, examining heat capacity is accurately outlined and furnishes significant evidence by analyzing Fig. \ref{figCc}. The first insight is taken by varying the parameter $Q$ and keeping the other parameters fixed. This configuration involves a set of sign-change phenomena, either by a continuous process (physical limitation point) or discontinuous behaviour (divergent point). The sole feature that emerges here is where the parameter $Q$ increases, the position of physical limitation points grows, and the set of divergent points is kept invariant. In addition, the system remains in a local thermal state because of the positivity pertinent to the heat capacity.
On the other hand, the variation of the coupling constant $c_q$, in contrast to the variation of the charge $Q$, does not lead to significant impacts on the physical limitation points, even as the parameter value of $c_q$ increases. This means that altering $c_q$ does not cause major changes in the critical points where physical properties of the smaller black holes might otherwise be expected to diverge or behave anomalously. Despite the changes in $c_q$, the number of divergent points remains fixed, indicating a stable framework for the system. Additionally, the thermal state of the black hole remains locally stable under variations in $c_q$, suggesting that the system can maintain thermal equilibrium without experiencing instability.

Moreover, the correction parameters $(\beta_1, \beta_2)$ have only minimal impacts on the behaviour of the heat capacity. These corrections introduce slight variations, which are more significant for smaller black holes compared to larger ones. For smaller black holes, the corrections might cause noticeable changes in the heat capacity, whereas for larger black holes, the effects are less pronounced. However, the crucial point to note is that these corrections due to thermal fluctuations do not alter the physical limitation points. They do not influence the locations where the system undergoes phase transitions or exhibits critical behaviour. Specifically, the corrections do not affect the second-order phase transition of the black hole, meaning the fundamental thermodynamic behaviour and stability characteristics remain unchanged. This highlights the robustness of the black hole's thermodynamic properties despite the presence of thermal fluctuations and parameter variations.

\section{Concluding Remarks}\label{sec7}
In this work, we have considered the corrected thermodynamics of non-linear magnetic charged AdS black holes by considering thermal fluctuations and calculating the corrected order terms of the thermodynamic potentials. Initially, we derived the expression for the entropy $S$ of the system and obtained a second-order corrected expression for the black hole's entropy. The entropy's first-order correction term depends on the entropy and temperature of the black hole. In contrast, the second-order corrected term depends inversely on the black hole's entropy. Subsequently, we obtained the corrected thermodynamic terms of the black hole, incorporating the second-order entropy correction. Our findings indicate that both the first and second-order correction terms have distinct impacts on the thermodynamic potentials of the black hole.

The variations in corrected entropy demonstrate that for small magnetic charge $Q$ and quintessence parameter $c_q$, the corrected entropy is a smooth increasing function of the horizon radius $r_{+}$. However, for larger $Q$, entropy shows a sudden rise, with peaks shifting towards larger radii. The quintessence parameter $c_q$ also influences entropy, leading to regular increases with significant peaks, except for $c_q=0.1$, where no significant peaks are observed. The corrected entropy remains positive across all cases for both first-order and second-order corrections.

The analysis of enthalpy variations shows that increasing $Q$ raises the enthalpy, particularly for smaller black holes, while $c_q$ reduces enthalpy, especially for larger black holes. Thermal fluctuations significantly increase enthalpy for smaller black holes and slightly decrease it for larger ones, indicating a shift in stability due to these fluctuations.

The Helmholtz free energy analysis reveals that an increase in $Q$ shifts the peak towards larger black holes and increases the area under the curve, while $c_q$ decreases the Helmholtz free energy. Correction parameters $\beta_1$ and $\beta_2$ exhibit distinct behaviours: $\beta_1$ initially increases the Helmholtz free energy for small black holes but decreases it beyond a certain radius, whereas $\beta_2$ gradually decreases it for small black holes.

The Gibbs free energy behaviour indicates that smaller black holes exhibit negative Gibbs free energy with increasing $Q$, signifying higher stability. The quintessence parameter $c_q$ decreases the Gibbs free energy peak without a significant horizontal shift. Correction parameters $\beta_1$ and $\beta_2$ also play crucial roles, with $\beta_1$ increasing and $\beta_2$ decreasing the Gibbs free energy initially but reversing their effects beyond a threshold radius.

Finally, our investigation into specific heat reveals critical insights into black hole stability. The specific heat calculations, influenced by thermal fluctuation parameters, indicate that large black holes with positive specific heat ensure stability against phase transitions. We identified conditions for physical limitation points and second-order phase transitions, independent of the correction parameters $\beta_1$ and $\beta_2$.

{Investigations associated with such types of black hole systems may be further extended to thermodynamical topologies, black hole chemistry, shadow behaviour, phase transitions etc. \cite{Mann:2024sru, Hazarika:2024imk, CamposDelgado:2023rti, Gogoi:2023wih, Upadhyay:2019hyw, Saghafi:2022pme,Ghaffarnejad:2022aqe} to understand the characteristics of such systems in more detail. }

Overall, our results show that the thermal fluctuations significantly impact the stability of small-sized black holes, making them less stable and resulting in a shorter lifetime. The first-order thermodynamic corrections tend to destabilize smaller black holes, while smaller values of second-order correction parameters also lead to instability. However, increasing the parameter values enhances the stability of smaller black holes. These findings provide valuable insights into the complex nature of black hole thermodynamics and the critical role of thermal fluctuations.

\section*{Acknowledgment}
 DJG acknowledges the contribution of the COST Action CA21136  -- ``Addressing observational tensions in cosmology with systematics and fundamental physics (CosmoVerse)". 

\section*{Declaration of competing interest}
The authors declare that they have no known competing financial interests or personal relationships that could have appeared to influence the work reported in this manuscript.

\section*{Data availability}
Data sharing does not apply to this article, as no data sets were analyzed or generated during the current study.


\begin{thebibliography}{00}

\bibitem{sb1} J. D. Bekenstein, ``Black Holes and the Second Law", \href{https://doi.org/10.1007/BF02757029}{Lett. Nuovo Cimento 4, 737 (1972).}

\bibitem{sb2} F. Belgiorno, S. L. Cacciatori, M. Clerici, V. Gorini, G. Ortenzi, L. Rizzi, E. Rubino, V. G. Sala, and D. Faccio, ``Hawking Radiation from Ultrashort Laser Pulse Filaments", \href{https://doi.org/10.1103/PhysRevLett.105.203901}{Phys. Rev. Lett. 105, 203901 (2010).}

\bibitem{li2} J. D. Bekenstein, ``Black Holes and Entropy", \href{https://doi.org/10.1103/PhysRevD.7.2333}{Phys. Rev. D 7, 2333 (1973).}

\bibitem{m1} J.M. Bardeen, B. Carter, and S.W. Hawking, ``The four laws of black hole mechanics", \href{https://doi.org/10.1007/BF01645742}{Commun. Math. Phys. 31, 161 (1973).}

\bibitem{m39} B. Pourhassan, M. Faizal and U. Debnath, “Effects of thermal fluctuations on the thermodynamics of modified Hayward black hole”, \href{https://doi.org/10.1140/epjc/s10052-016-3998-8}{Eur. Phys. J. C 76, 145 (2016).}

\bibitem{m13} N. Islam, P.A. Ganai, and S. Upadhyay, “Thermal fluctuations to the thermodynamics of a non-rotating BTZ black hole”, \href{https://doi.org/10.1093/ptep/ptz113}{Prog. Theor. Exp. Phys. 103B06, 1 (2019).}

\bibitem{m14} S. Upadhyay, “Leading-order corrections to charged rotating AdS black holes thermodynamics”, \textcolor{blue}{Gen Relativ Gravit 50, 128 (2018).}

\bibitem{m15} S. Upadhyay and B. Pourhassan, “Logarithmic-corrected van der Waals black holes in higher-dimensional AdS space”, \textcolor{blue}{Prog. Theor. Exp. Phys. 013B03, 1 (2019).}

\bibitem{m15a} Wald, R.M. The Thermodynamics of Black Holes. \textcolor{blue}{Living Rev. in Rel. 4, 6 (2001).}

\bibitem{m16} S. Upadhyay, S.H. Hendi, S. Panahiyan, and B.E. Panah, “Thermal fluctuations of charged black holes in gravity's rainbow”, \textcolor{blue}{Prog. Theor. Exp. Phys. 093E01, 1 (2018).}

\bibitem{m17} B. Pourhassan, S. Upadhyay, H. Saadat, and H. Farahani, “Quantum gravity effects on HoravaLifshitz black hole”, \textcolor{blue}{Nucl. Phys. B 928, 415 (2018).}

\bibitem{m18} S. Upadhyay, “Quantum corrections to thermodynamics of quasi topological black holes”, \textcolor{blue}{Phys. Lett. B 775, 130 (2017).}

\bibitem{m19}  B. Pourhassan, M. Faizal, S. Upadhyay and L.A. Asfar, “Thermal fluctuations in a hyperscaling-violation background”, \textcolor{blue}{Eur. Phys. J. C 77, 555 (2017).}

\bibitem{m20}  S. Upadhyay, S. Soroushfar and R. Saffari, “Perturbed thermodynamics and thermodynamic geometry of a static black hole in $f(R)$ gravity”, \textcolor{blue}{Mod. Phys. Lett. A 36, 2150212 (2021).}

\bibitem{li3} B. Pourhassan, H. Farahani, and S. Upadhyay, Thermodynamics of Higher Order Entropy Corrected Schwarzschild-Beltrami-de Sitter Black Hole.

\bibitem{m25} J. Jing and M.L. Yan, “Statistical Entropy of a Stationary Dilaton Black Hole from Cardy Formula”, \textcolor{blue}{Phys. Rev. D 63, 024003 (2001).}

\bibitem{m26} B. Pourhassan, S. Upadhyay and H. Farahani, “Thermodynamics of Higher Order Entropy Corrected Schwarzschild-Beltrami-de Sitter Black Hole”, \textcolor{blue}{Int. J. Mod. Phys. A 34, 1950158 (2019).}

\bibitem{m28} A. Chamblin, R. Emparan, C. Johnson and R. Myers, “Holography, thermodynamics, and fluctuations of charged AdS black holes”, \textcolor{blue}{Phys. Rev. D 60, 104026 (1999).}

\bibitem{m29} S.W. Hawking and D.N. Page, “Thermodynamics of black holes in anti-de Sitter space”, \textcolor{blue}{Commun. Math. Phys. 87, 577 (1983).}

\bibitem{m36} B. Pourhassan, M. Faizal and S. Capozziello, “Testing quantum gravity through dumb holes”, \textcolor{blue}{Annals Phys. 377, 108 (2017).}

\bibitem{m37} B. Pourhassan and M. Faizal, “Effect of thermal fluctuations on a charged dilatonic black Saturn”, \textcolor{blue}{Phys. Lett. B 755, 444 (2016).}

\bibitem{m38} B. Pourhassan and M. Faizal, “Thermodynamics of a sufficient small singly spinning Kerr-AdS black hole”, \textcolor{blue}{Nucl. Phys. B 913, 834 (2016).}

\bibitem{Banerjee1}
R.~Banerjee and B.~R.~Majhi, ``Quantum Tunneling Beyond Semiclassical Approximation'',
\textcolor{blue}{JHEP 06, 095 (2008).}

\bibitem{Banerjee2}
R.~Banerjee and B.~R.~Majhi,
``Quantum Tunneling, Trace Anomaly and Effective Metric'',
\textcolor{blue}{Phys. Lett. B 674, 218-222 (2009).}

\bibitem{Lambiase01} G. Lambiase et al., ``Investigating the Connection between Generalized Uncertainty Principle and Asymptotically Safe Gravity in Black Hole Signatures through Shadow and Quasinormal Modes", \textcolor{blue}{Eur. Phys. J. C 83, 679 (2023).}

\bibitem{dj}D. J. Gogoi et al., ``Joule-Thomson Expansion and Optical Behaviour of Reissner-Nordström-Anti-de Sitter Black Holes in Rastall Gravity Surrounded by a Quintessence Field", \textcolor{blue}{Fortschritte Der Physik 71, 2300010 (2023).}

\bibitem{Banerjee3}
R.~Banerjee and S.~K.~Modak,
``Exact Differential and Corrected Area Law for Stationary Black Holes in Tunneling Method'',
\textcolor{blue}{JHEP 05, 063 (2009).}

\bibitem{Banerjee4}
R.~Banerjee, S.~K.~Modak and S.~Samanta,
``Second Order Phase Transition and Thermodynamic Geometry in Kerr-AdS Black Hole'',
\textcolor{blue}{Phys. Rev. D 84, 064024 (2011).}


\bibitem{Banerjee5}
R.~Banerjee, S.~Ghosh and D.~Roychowdhury,
``New type of phase transition in Reissner Nordstr\"om\textendash{}AdS black hole and its thermodynamic geometry'',
\textcolor{blue}{Phys. Lett. B 696, 156-162 (2011).}

\bibitem{mm1} S. Upadhyay, B. Pourhassan and H. Farahani, “P–V criticality of first-order entropy corrected AdS black holes in massive gravity”, \textcolor{blue}{Phys. Rev. D 95, 106014 (2017).}

\bibitem{mm2}  D.~J.~Gogoi and S.~Ponglertsakul,
``Constraints on quasinormal modes from black hole shadows in regular non-minimal Einstein Yang\textendash{}Mills gravity,''
Eur. Phys. J. C \textbf{84}, no.6, 652 (2024)
doi:10.1140/epjc/s10052-024-12946-9
[arXiv:2402.06186 [gr-qc]].



\bibitem{mm3} A. Pourdarvish, J. Sadeghi, H. Farahani, and B. Pourhassan, “Thermodynamics and Statistics of Goedel Black Hole with Logarithmic Correction”, \textcolor{blue}{Int. J. Theor. Phys. 52, 3560 (2013).}


\bibitem{mm4} T.R. Govindarajan, R.K. Kaul, and V. Suneeta, “Logarithmic correction to the Bekenstein-Hawking entropy of the BTZ black hole”, \textcolor{blue}{Class. Quantum Grav. 18, 2877 (2001).}


\bibitem{mm5} D.~J.~Gogoi, A.~\"Ovg\"un and D.~Demir,
``Quasinormal modes and greybody factors of symmergent black hole,''
Phys. Dark Univ. \textbf{42}, 101314 (2023)
doi:10.1016/j.dark.2023.101314
[arXiv:2306.09231 [gr-qc]].


\bibitem{mm6} M. Cvetic and S.S. Gubser, “Phases of R-charged Black Holes, Spinning Branes and Strongly Coupled Gauge Theories”, \textcolor{blue}{JHEP 9904, 024 (1999).}

\bibitem{mm7} B. Pourhassan, M. Faizal and S. Ahmad Ketabi, “Logarithmic correction of the BTZ black hole and adaptive model of graphene”, \textcolor{blue}{Int. J. Mod. Phys. D 27, 1850118 (2018).}

\bibitem{mm8} B. Pourhassan, M. Faizal, Z. Zaz and A. Bhat, “Quantum fluctuations of a BTZ black hole in massive gravity”, \textcolor{blue}{Phys. Lett. B 773, 325 (2017).}

\bibitem{mm9} D.~J.~Gogoi, N.~Heidari, J.~Kr\'iz and H.~Hassanabadi,
``Quasinormal Modes and Greybody Factors of de Sitter Black Holes Surrounded by Quintessence in Rastall Gravity,''
Fortsch. Phys. \textbf{72}, no.3, 2300245 (2024)
doi:10.1002/prop.202300245
[arXiv:2307.09976 [gr-qc]].



\bibitem{mm10} B. Pourhassan, M. Faizal, “The lower bound violation of shear viscosity to entropy ratio due to logarithmic correction in STU model”, \textcolor{blue}{Eur. Phys. J. C 77, 96 (2017).}


\bibitem{mm11}  M. Faizal, A. Ashour, M. Alcheikh, L.-Al Asfar, S. Alsaleh, and A. Mahroussah, “Quantum fluctuations from thermal fluctuations in Jacobson formalism”, \textcolor{blue}{Eur. Phys. J. C 77, 608 (2017).}


\bibitem{mm12} S. Das, P. Majumdar, and R.K. Bhaduri, “General Logarithmic Corrections to Black Hole Entropy”, \textcolor{blue}{Class. Quantum Grav. 19, 2355 (2002).}

\bibitem{mm13}B. Pourhassan and M. Faizal, ``Thermal Fluctuations in a Charged AdS Black Hole", \textcolor{blue}{EPL 111, 40006 (2015).}


\bibitem{mm14}
S. Vagnozzi et al., ``Horizon-scale tests of gravity theories and fundamental physics from the Event Horizon Telescope image of Sagittarius A*'', \textcolor{blue}{Class. Quantum Grav. 40, 165007 (2023).}


\bibitem{mm15}
G.~Lambiase, D.~J.~Gogoi, R.~C.~Pantig and A.~\"Ovg\"un,
``Shadow and quasinormal modes of the rotating Einstein-Euler-Heisenberg black holes,''
[arXiv:2406.18300 [gr-qc]].

\bibitem{mm16}
D.~J.~Gogoi,
``Violation of Hod\textquoteright{}s conjecture and probing it with optical properties of a 5-D black hole in Einstein Gauss\textendash{}Bonnet Bumblebee theory of gravity,''
Phys. Dark Univ. \textbf{45}, 101535 (2024)
doi:10.1016/j.dark.2024.101535
[arXiv:2405.02455 [gr-qc]].

\bibitem{mm17}
P.~Bhar, D.~J.~Gogoi and S.~Ponglertsakul,
``Noncommutative black hole in de Rham-Gabadadze-Tolley like massive gravity,''
[arXiv:2404.10627 [gr-qc]].


\bibitem{mm18}
Y.~Sekhmani, D.~J.~Gogoi, R.~Myrzakulov and J.~Rayimbaev,
``Phase structures and critical behavior of rational non-linear electrodynamics Anti de Sitter~black holes in Rastall gravity,''
Commun. Theor. Phys. \textbf{76}, no.4, 045403 (2024)
doi:10.1088/1572-9494/ad30f4
[arXiv:2403.04888 [gr-qc]].

\bibitem{mm19}
N. Islam and P.A. Ganai, “Quantum corrections to AdS black hole in massive gravity”, \textcolor{blue}{Int. J. Mod. Phys. A 34, 1950225 (2019).}


\bibitem{mm20}
Y.~Sekhmani, J.~Rayimbaev, G.~G.~Luciano, R.~Myrzakulov and D.~J.~Gogoi,
``Phase structure of charged AdS black holes surrounded by exotic fluid with modified Chaplygin equation of state,''
Eur. Phys. J. C \textbf{84}, no.3, 227 (2024)
doi:10.1140/epjc/s10052-024-12597-w
[arXiv:2311.02448 [gr-qc]].

\bibitem{mm21}
Y.~Sekhmani, D.~J.~Gogoi, M.~Baouahi and I.~Dahiri,
``Thermodynamic geometry of STU black holes,''
Phys. Scripta \textbf{98}, no.10, 105014 (2023)
doi:10.1088/1402-4896/acf7fb

\bibitem{mm22}
S. Chougule, S. Dey, B. Pourhassan and M. Faizal, “BTZ black holes in massive gravity”, \textcolor{blue}{Eur. Phys. J. C 78, 685 (2018).}

\bibitem{mm23}
D.~J.~Gogoi, J.~Bora, M.~Koussour and Y.~Sekhmani,
``Quasinormal modes and optical properties of 4-D black holes in Einstein Power-Yang\textendash{}Mills gravity,''
Annals Phys. \textbf{458}, 169447 (2023)
doi:10.1016/j.aop.2023.169447
[arXiv:2306.14273 [gr-qc]].


\bibitem{mm24}
M. Cvetic and S.S. Gubser, “Thermodynamic Stability and Phases of General Spinning Branes”, \textcolor{blue}{JHEP 9907, 010 (1999).}



\bibitem{kruglov}S. I. Kruglov, ``Magnetic Black Holes in AdS Space with Nonlinear Electrodynamics, Extended Phase Space Thermodynamics and Joule–Thomson Expansion", \href{https://doi.org/10.1142/S0219887823500081}{Int. J. Geom. Methods Mod. Phys. 20, 2350008 (2023).}

\bibitem{nam} Cao H Nam. ``On non-linear magnetic-charged black hole surrounded by quintessence" \href{https://doi.org/10.1007/s10714-018-2380-6}{Gen. Rel. Grav., 50, 6, 57 (2018).}


\bibitem{ndo} R. Ndongmo, S. Mahamat, C. B. Tabi, T. B. Bouetou, and T. C. Kofane, ``Thermodynamics of Non-Linear Magnetic-Charged AdS Black Hole Surrounded by Quintessence, in the Background of Perfect Fluid Dark Matter", \href{https://doi.org/10.1016/j.dark.2023.101299}{Physics of the Dark Universe 42, 101299 (2023).}

\bibitem{Jawad:2020ihz}
A.~Jawad,
``Consequences of Thermal Fluctuations of Well-Known Black Holes in Modified Gravity,''
Class. Quant. Grav. \textbf{37}, no.18, 185020 (2020)
doi:10.1088/1361-6382/ab9ad5
[arXiv:2008.11033 [gr-qc]].

\bibitem{Zhang:2018nep}
M.~Zhang,
``Corrected thermodynamics and geometrothermodynamics for anti-de Sitter black hole,''
Nucl. Phys. B \textbf{935}, 170-182 (2018)
doi:10.1016/j.nuclphysb.2018.08.010

\bibitem{Pourhassan:2017bip}
B.~Pourhassan and K.~Kokabi,
``Effects of higher-order corrected entropy on the black hole physics,''
Can. J. Phys. \textbf{96}, no.3, 262-267 (2018)
doi:10.1139/cjp-2017-0550

\bibitem{Jawad:2017mwt}
A.~Jawad and M.~U.~Shahzad,
``Effects of Thermal Fluctuations on Non-minimal Regular Magnetic Black Hole,''
Eur. Phys. J. C \textbf{77}, no.5, 349 (2017)
doi:10.1140/epjc/s10052-017-4914-6
[arXiv:1705.10012 [gr-qc]].

\bibitem{Pourhassan:2016qoz}
B.~Pourhassan, M.~Faizal and U.~Debnath,
``Effects of Thermal Fluctuations on the Thermodynamics of Modified Hayward Black Hole,''
Eur. Phys. J. C \textbf{76}, no.3, 145 (2016)
doi:10.1140/epjc/s10052-016-3998-8
[arXiv:1603.01457 [gr-qc]].

\bibitem{Pourhassan:2020yei}
B.~Pourhassan,
``Exponential corrected thermodynamics of black holes,''
J. Stat. Mech. \textbf{2107}, 073102 (2021)
doi:10.1088/1742-5468/ac0f6a
[arXiv:2010.03946 [gr-qc]].

\bibitem{Chatterjee:2020iuf}
A.~Chatterjee and A.~Ghosh,
``Exponential Corrections to Black Hole Entropy,''
Phys. Rev. Lett. \textbf{125}, no.4, 041302 (2020)
doi:10.1103/PhysRevLett.125.041302
[arXiv:2007.15401 [gr-qc]].

\bibitem{Pourhassan:2022cvn}
B.~Pourhassan, M.~Dehghani, S.~Upadhyay, I.~Sakalli and D.~V.~Singh,
``Exponential corrected thermodynamics of Born\textendash{}Infeld BTZ black holes in massive gravity,''
Mod. Phys. Lett. A \textbf{37}, no.33n34, 2250230 (2022)
doi:10.1142/S0217732322502303
[arXiv:2301.01603 [gr-qc]].

\bibitem{Nadeem-ul-islam:2020dhd}
Nadeem-ul-islam and P.~A.~Ganai,
``First-order corrected thermodynamic potentials characterizing BTZ black hole in massive gravity,''
Int. J. Mod. Phys. A \textbf{35}, no.18, 2050080 (2020)
doi:10.1142/S0217751X20500803

\bibitem{Fatima:2024gji}
G.~Fatima, S.~Shaukat, F.~Javed and G.~Mustafa,
``Greybody factors, quasi-normal modes and thermal fluctuations of quantum-corrected Schwarzschild black hole surrounded by quintessence,''
Phys. Dark Univ. \textbf{45}, 101521 (2024)
doi:10.1016/j.dark.2024.101521

\bibitem{Pourhassan:2018scc}
B.~Pourhassan, K.~Kokabi and Z.~Sabery,
``Higher order corrected thermodynamics and statistics of Kerr\textendash{}Newman\textendash{}G\"odel black hole,''
Annals Phys. \textbf{399}, 181-192 (2018)
doi:10.1016/j.aop.2018.10.011
[arXiv:1811.03152 [gr-qc]].


\bibitem{li2012galactic} M-H. Li and K.-C. Yang ``Galactic dark matter in the phantom field"
\newblock {\em Physical Review D}, 86(12):123015, 2012.

\bibitem{xu2018kerr}
Zhaoyi Xu, Xian Hou, and Jiancheng Wang.
\newblock Kerr-anti-de sitter/de sitter black hole in perfect fluid dark matter
  background.
\newblock {\em Classical and Quantum Gravity}, 35(11):115003, 2018.

\bibitem{xu2019perfect}
Zhaoyi Xu, Xian Hou, Jiancheng Wang, and Yi~Liao.
\newblock Perfect fluid dark matter influence on thermodynamics and phase
  transition for a reissner-nordstrom-anti-de sitter black hole.
\newblock {\em Advances in High Energy Physics}, 2019, 2019.

\bibitem{sadeghi2020universal}
J~Sadeghi, E~Naghd Mezerji, and S~Noori Gashti.
\newblock Universal relations and weak gravity conjecture of ads black holes
  surrounded by perfect fluid dark matter with small correction.
\newblock {\em arXiv preprint arXiv:2011.14366}, 2020.

\bibitem{salazar1987duality}
Humberto Salazar~I, Alberto Garc{\'\i}a~D, and Jerzy Pleba{\'n}ski.
\newblock Duality rotations and type d solutions to einstein equations with
  nonlinear electromagnetic sources.
\newblock {\em Journal of mathematical physics}, 28(9):2171--2181, 1987.

\bibitem{novello2000singularities}
M~Novello, SE~Perez Bergliaffa, and JM~Salim.
\newblock Singularities in general relativity coupled to nonlinear
  electrodynamics.
\newblock {\em Classical and Quantum Gravity}, 17(18):3821, 2000.

\bibitem{nam2020higher}
Cao~H Nam.
\newblock Higher dimensional charged black hole surrounded by quintessence in
  massive gravity.
\newblock {\em General Relativity and Gravitation}, 52(1):1, 2020.

\bibitem{nam2018on}
Cao~H Nam.
\newblock On non-linear magnetic-charged black hole surrounded by quintessence.
\newblock {\em General Relativity and Gravitation}, 50(6):57, 2018.

\bibitem{nam2018non}
Cao~H Nam.
\newblock Non-linear charged ds black hole and its thermodynamics and phase
  transitions.
\newblock {\em The European Physical Journal C}, 78(5):418, 2018.

\bibitem{chen2020optical}
Yuan Chen, He-Xu Zhang, Tian-Chi Ma, and Jian-Bo Deng.
\newblock Optical properties of a nonlinear magnetic charged rotating black
  hole surrounded by quintessence with a cosmological constant.
\newblock {\em arXiv preprint arXiv:2009.03778}, 2020.

\bibitem{ma2021shadow}
Tian-Chi Ma, He-Xu Zhang, Peng-Zhang He, Hao-Ran Zhang, Yuan Chen, and Jian-Bo Deng.

\newblock Shadow cast by a rotating and nonlinear magnetic-charged black hole
  in perfect fluid dark matter.
\newblock {\em Modern Physics Letters A}, page 2150112, 2021.

\bibitem{sadeghi2020ads}
M. Sadeghi, (2020). AdS black brane solution surrounded by quintessence in massive gravity and KSS bound. Modern Physics Letters A.

\bibitem{ghosh2018lovelock}
Ghosh, S.G., Maharaj, S.D., Baboolal, D. et al. Lovelock black holes surrounded by quintessence. Eur. Phys. J. C 78, 90 (2018). https://doi.org/10.1140/epjc/s10052-018-5570-1

\bibitem{bohmer2015interacting}
Boehmer, C. G., Tamanini, N., \& Wright, M. (2015). Interacting quintessence from a variational approach Part I: Algebraic couplings. ArXiv. https://doi.org/10.1103/PhysRevD.91.123002

\bibitem{gonzalez2008exact}
A NEW EXACT STATIC THIN DISK WITH A CENTRAL BLACK HOLE
GUILLERMO A. GONZÁLEZ (Colombia)
The Eleventh Marcel Grossmann Meeting. September 2008, 2325-2327

\bibitem{rizwan2020coexistent}
C. L. A. Rizwan, A. N. Kumara, K. Hegde, and D. Vaid, {\em Coexistent Physics and Microstructure of the Regular Bardeen Black Hole in Anti-de Sitter Spacetime}, \href{https://doi.org/10.1016/j.aop.2020.168320}{Annals of Physics 422, 168320 (2020)}.

\bibitem{zhang2021regular}
H.~X.~Zhang, Y.~Chen, T.~C.~Ma, P.~Z.~He and J.~B.~Deng,
``Bardeen black hole surrounded by perfect fluid dark matter,''
Chin. Phys. C \textbf{45}, no.5, 055103 (2021)
doi:10.1088/1674-1137/abe84c
[arXiv:2007.09408 [gr-qc]].

\bibitem{Kiselev2003}
V. V. Kiselev, {\em Quintessence and Black Holes}, \href{https://doi.org/10.1088/0264-9381/20/6/310}{Class. Quantum Grav. 20, 1187 (2003).}

\bibitem{Norte:2024crb}
R.~A.~Norte,
``Negative-temperature pressure in black holes,''
EPL \textbf{145}, no.2, 29001 (2024)
doi:10.1209/0295-5075/ad2088
[arXiv:2402.18570 [gr-qc]].

\bibitem{Toledo:2019amt}
J.~M.~Toledo and V.~B.~Bezerra,
``Some remarks on the thermodynamics of charged AdS black holes with cloud of strings and quintessence,''
Eur. Phys. J. C \textbf{79}, no.2, 110 (2019)
doi:10.1140/epjc/s10052-019-6616-8



\bibitem{hawp} 
S.~W.~Hawking and D.~N.~Page,
``Thermodynamics of Black Holes in anti-De Sitter Space,''
Commun. Math. Phys. \textbf{87}, 577 (1983)

\bibitem{More:2004hv}
S.~S.~More,
``Higher order corrections to black hole entropy,''
Class. Quant. Grav. \textbf{22}, 4129-4140 (2005)
[arXiv:gr-qc/0410071 [gr-qc]].

\bibitem{PF}
B.~Pourhassan and M.~Faizal,
``Thermodynamics of a sufficient small singly spinning Kerr-AdS black hole,''
Nucl. Phys. B \textbf{913}, 834-851 (2016)
[arXiv:1611.00131 [gr-qc]].

\bibitem{SPR}
J.~Sadeghi, B.~Pourhassan and M.~Rostami,
``P-V criticality of logarithm-corrected dyonic charged AdS black holes,''
Phys. Rev. D \textbf{94}, no.6, 064006 (2016)
doi:10.1103/PhysRevD.94.064006
[arXiv:1605.03458 [gr-qc]].

\bibitem{Mandal:2023ahb}
S.~Mandal, S.~Das, D.~J.~Gogoi and A.~Pramanik,
``Leading-order corrections to the thermodynamics of Rindler modified Schwarzschild black hole,''
Phys. Dark Univ. \textbf{42}, 101349 (2023)
doi:10.1016/j.dark.2023.101349
[arXiv:2308.05712 [gr-qc]].

\bibitem{Sahabandu:2005ma}
C.~Sahabandu, P.~Suranyi, C.~Vaz and L.~C.~R.~Wijewardhana,
``Thermodynamics of static black objects in D dimensional Einstein-Gauss-Bonnet gravity with D-4 compact dimensions,''
Phys. Rev. D \textbf{73} (2006), 044009
doi:10.1103/PhysRevD.73.044009
[arXiv:gr-qc/0509102 [gr-qc]].

\bibitem{Cai:2003kt}
R.~G.~Cai,
``A Note on thermodynamics of black holes in Lovelock gravity,''
Phys. Lett. B \textbf{582} (2004), 237-242
doi:10.1016/j.physletb.2004.01.015
[arXiv:hep-th/0311240 [hep-th]].


\bibitem{Mann:2024sru}
R.~B.~Mann,
``Recent Developments in Holographic Black Hole Chemistry,''
JHAP \textbf{4}, no.1, 1-26 (2024)
doi:10.22128/jhap.2023.757.1067
[arXiv:2403.02864 [hep-th]].

\bibitem{Hazarika:2024imk}
B.~Hazarika, N.~J.~Gogoi and P.~Phukon,
``Revisiting thermodynamic topology of Hawking-Page and Davies type phase transitions,''
[arXiv:2404.02526 [hep-th]].

\bibitem{CamposDelgado:2023rti}
R.~Campos Delgado,
``Quantum gravitational corrected evolution equations of charged black holes,''
JHAP \textbf{3}, no.1, 39-48 (2023)
doi:10.22128/jhap.2023.643.1037
[arXiv:2302.07835 [gr-qc]].

\bibitem{Gogoi:2023wih}
N.~J.~Gogoi and P.~Phukon,
``Thermodynamic topology of 4D Euler\textendash{}Heisenberg-AdS black hole in different ensembles,''
Phys. Dark Univ. \textbf{44}, 101456 (2024)
doi:10.1016/j.dark.2024.101456
[arXiv:2312.13577 [hep-th]].

\bibitem{Upadhyay:2019hyw}
S.~Upadhyay, Nadeem-ul-islam and P.~A.~Ganai,
``A modified thermodynamics of rotating and charged BTZ black hole,''
JHAP \textbf{2}, no.1, 25-48 (2022)
doi:10.22128/jhap.2021.454.1004
[arXiv:1912.00767 [gr-qc]].

\bibitem{Saghafi:2022pme}
S.~Saghafi and K.~Nozari,
``Shadow behavior of the quantum-corrected Schwarzschild black hole immersed in holographic quintessence,''
JHAP \textbf{3}, no.1, 31-38 (2022)
doi:10.22128/jhap.2022.515.1019
[arXiv:2306.13767 [hep-th]].

\bibitem{Ghaffarnejad:2022aqe}
H.~Ghaffarnejad and E.~Ghasemi,
``Magnetic charge effects on thermodynamic phase transition of modified Anti de Sitter Ay\'on-Beato-Garc\'\i{}a black holes with five parameters,''
JHAP \textbf{3}, no.1, 47-56 (2022)
doi:10.22128/jhap.2022.524.1022
[arXiv:2204.02979 [gr-qc]].


\end{thebibliography}
\end{document}